\documentclass{article}
\usepackage[a4paper, total={7in, 9in}]{geometry}
\usepackage{hyperref}
\hypersetup{
    colorlinks=true,
    linkcolor=red,
    filecolor=magenta,      
    urlcolor=red,
}
\usepackage{amsmath,amssymb,amsfonts,amstext}
\usepackage{graphicx}
\usepackage{lipsum}
\usepackage{float}
\usepackage{amsthm}
\usepackage{xcolor}
\usepackage[normalem]{ulem}
\newtheorem{theorem}{Theorem}

\newtheorem{corollary}{Corollary}
\newtheorem{remark}{Remark}
\newtheorem{property}{Property}
\usepackage{algorithm}

\usepackage[style=ieee]{biblatex}
\usepackage{lipsum,graphicx}

\usepackage{hologo}

\title{\textbf{Point process simulation of generalised inverse Gaussian processes and estimation of the Jaeger integral}}

\author{\textbf{Simon~Godsill and Yaman~K{\i}ndap} \\
        {\small Signal Processing and Communications Laboratory} \\
        {\small University of Cambridge, UK}
       }
       
\date{\today}

\begin{document}
\maketitle

\begin{abstract}
In this paper novel simulation methods are provided for the generalised inverse Gaussian (GIG) L\'{e}vy process. Such processes are intractable for simulation except in certain special edge cases, since the L\'{e}vy density associated with the GIG process is expressed as  an integral involving certain Bessel Functions, known as the Jaeger integral in diffusive transport applications. We here show for the first time how to solve the problem indirectly, using generalised shot-noise methods to simulate the underlying point processes and constructing an auxiliary variables approach that avoids any direct calculation of the integrals involved. The resulting augmented bivariate process is still intractable and so we propose a novel thinning method based on upper bounds on the intractable integrand. Moreover our approach leads to lower and upper bounds on the Jaeger integral itself, which may be compared with other approximation methods. The shot noise method involves a truncated infinite series of decreasing random variables, and as such is approximate, although the series are found to be rapidly convergent in most cases. We note that the GIG process is the required Brownian motion subordinator for the generalised hyperbolic (GH) L\'{e}vy process and so our simulation approach will straightforwardly extend also to the simulation of these intractable proceses. Our new methods will find application in forward simulation of processes of GIG and GH type, in financial and engineering data, for example, as well as inference for states and parameters of stochastic processes driven by GIG and GH L\'{e}vy processes.

{\textbf{Keywords:} L\'{e}vy process, generalised inverse Gaussian process, generalised hyperbolic process, diffusive transport, Jaeger integral, rejection sampling, Monte Carlo methods, Poisson point process, thinning}
\end{abstract}

\section{Introduction}

The L\'{e}vy process is a fundamental tool for the study of continuous time stochastic phenomena \cite{Bertoin_1997,Barndorff-NielsenMikoschResnick2001,applebaum_2009}. The most familiar example is of course Brownian motion, but it is well known that the Gaussian assumption is inadequate for modelling of real-world phenomena, which are often more heavy-tailed than the Gaussian. Applications are found in areas such as financial modelling \cite{Mandelbrot1963,Fama1965a,Cont:2003}, communications \cite{Azzaoui2010,Fahs2012,FreitasEganClavierEtAl2017,LiebeherrBurchardCiucu2012,ShevlyakovKim2006,WarrenThomas1991},
signal processing \cite{Nikias1995}, image analysis \cite{Achim2001,Achim2006}
and audio processing \cite{Godsill_Rayner_1996,Lombardi2006}. 
Non-Gaussian heavy-tailed effects are also important in the climatological sciences 
\cite{Katz1992,Katz2002}, in the medical sciences
\cite{ChenWangMcKeown2010} and for the understanding of sparse modelling/Compressive Sensing \cite{UnserTaftiAminiEtAl2014,UnserTaftiSun2014,Unser2014,AminiUnser2014,CarrilloRamirezArceEtAl2016,Lopes2016,ZhouYu2017,Tzagkarakis2009,AchimBuxtonTzagkarakisEtAl2010}.

In this paper we study a very general class of non-Gaussian L\'{e}vy processes, the generalised inverse Gaussian process \cite{Barndorff-NielsenMikoschResnick2001,Eberlein2003}, a process which can capture various degrees of heavy-tailed behaviour, including the gamma process, the inverse Gaussian and the reciprocal-gamma process, as well as processes that lie somewhere `in between' these edge cases \cite{Eberlein2003}. Our work also enables direct simulation of the mean-variance mixtures of GIG processes, leading to the generalised hyperbolic (GH) processes \cite{Eberlein_2001}. Important sub-classes of the GH process include the asymmetric Student-t process which has previously been intractable for simulation and inference to our knowledge (this class is entirely distinct from the Student-t processes developed in the Machine Learning literature \cite{shah_2014}).  The use of asymmetric Student-t models in financial econometric applications are discussed in \cite{ZhuGalbraith2010}.

Rosinski \cite{Rosinski_2001} presents a generalised shot-noise representation of non-Gaussian L\'{e}vy processes, and it is this approach that we develop here for the GIG and GH processes. Our previous work using the shot noise representation has focussed on stable law processes and their applications in engineering, see \cite{Lemke_Godsill_2015,Riabiz2020,Godsill_Riabiz_Kont_2019} and references therein. There have been relevant developments in various special cases over recent years, including 
\cite{B_N_1997} who present the theory of normal-inverse Gaussian (NIG) processes, 
\cite{Rydberg_1997} who present approximate sampling methods for the  NIG  case, while
\cite{B_N_Shephard_2001} give applications of series based methods for non-Gaussian Ornstein-Uhlenbeck (OU) processes. 
\cite{Zhang_2011} provided rejection sampling and series based simulations for the  GIG OU process using the concepts of \cite{Rosinski_2001}, but these are applied to a different Background driving L\'{e}vy process (BDLP) than our work and indeed require evaluation of a generally intractable integral involving Bessel Functions. In addition \cite{Zhang_2011a,Qu_etal_2020,Grabchak2019RejectionSF} have provided exact simulation methods for the related class of tempered stable (TS) processes,  while recent relevant literature on GIG L\'{e}vy fields can be found in \cite{barth2017approximation}. It should be noted that the shot noise method involves infinite series of decreasing random variables, and in practice the series must be truncated at some finite limit in simulation. This truncation is the only approximation involved in our methods, which are otherwise exact, and in most cases the series are found to converge rapidly as the  number of terms increases.


The distribution of the GIG L\'{e}vy process at $t=1$, the GIG distribution, possesses
a three parameter probability density function defined for positive real random variables as follows \cite{Eberlein2003}:

\begin{equation}
    f_{GIG}(x) = \left( \frac{\gamma}{\delta} \right)^{\lambda} \frac{1}{2 K_{\lambda}(\delta \gamma)} x^{\lambda-1} e^{-\frac{1}{2} \left( \delta^2 x^{-1} + \gamma^2x \right)} \mathcal{I}_{x > 0}
\label{eqn:GIG_pdf}
\end{equation}

\noindent where $\lambda \in \mathbb{R}$, $K_{\lambda}(\cdot)$ is the modified Bessel function of the second kind and $\cal I$ is the indicator function. The support of parameters $\gamma$ and $\delta$ depend on the sign of $\lambda$ such that:

\begin{align*}
    &\delta \geq 0, \quad \gamma > 0, \quad \text{if} \, \lambda > 0, \\
    &\delta > 0, \quad \gamma > 0, \quad \text{if} \, \lambda = 0, \\
    &\delta > 0, \quad \gamma \geq 0, \quad \text{if} \, \lambda < 0.
\end{align*}

\noindent as discussed in \cite{Eberlein2003}. Random variate generation algorithms for a GIG variable are studied in \cite{Devroye_2014,Hormann2013}.

It is shown in \cite{Barndorffnielsen1977InfiniteDO} that the GIG distribution is infinitely divisible and hence can be the distribution of a L\'{e}vy process at time $t=1$. Furthermore, particular values of the parameters lead to special cases of the GIG distribution such as the inverse Gaussian $(\lambda = -1/2)$, Gamma ($\delta=0$, $\lambda>0$) and the reciprocal-Gamma ($\gamma = 0$, $\lambda<0$) distributions, and in these limit cases the normalising constant is replaced by the normalising constant of these well known distributions. 

The principal contribution of this paper is to provide a comprehensive suite of methods for simulation of GIG processes, without the need for evaluation of intractable integrals, beyond pointwise evaluation of the relevant Bessel functions.  An auxiliary variables approach transforms the univariate GIG point process into a bivariate point process having the GIG process as its marginal by construction, and requiring no explicit evaluation of integrals. We derive tractable dominating measures for the augmented GIG L\'{e}vy density, hence leading to a random thinning methodology for generation of jumps of the underlying marginal  GIG process. The whole procedure is carried out through generation of familiar random variables (from the gamma family) and point processes (both gamma and tempered stable). In addition we are able to bound the average acceptance rates of the random variate generation and also to provide upper and lower bounds on the  GIG L\'{e}vy density and the corresponding Jaeger integral. Finally the whole methodology is made accessible to practitioners through the publication of Matlab and Python \href{https://github.com/yamankindap/GiG}{code}\footnote{Matlab and Python code can be found in: https://github.com/yamankindap/GiG} to implement the GIG sampling schemes. 

Section 2 presents the necessary preliminaries for simulation of L\'{e}vy processes and their corresponding point processes, using a generalised shot-noise approach. Section 3 gives the specific forms for the GIG L\'{e}vy density and derives bounds on these densities, as well as a generic thinning method for tractable sampling of the underlying point processes, and presents in detail the simulation method for two different parameter ranges of the process. Section 4 presents example simulations, compared with exact simulations of the GIG random variable and finally Section 5 discusses the application of GIG processes in simulation and inference for more general processes including the generalised hyperbolic process.

\section{Shot noise representations, L\'{e}vy densities and thinning}

In this section we present the required preliminaries about L\'{e}vy processes and shot-noise simulation of those processes, using the framework of \cite{Rosinski_2001}. The characteristic function for a general non-negative-valued  L\'{e}vy process $W(t)$ having no drift or Brownian motion part is given by \cite{Kallenberg_2002}, Corollary 15.8, as:

\begin{equation}
    E \left[ \exp(iuW(t)) \right] = \exp \left( t \left[\int_{(0,\infty)} (e^{iuw} -1)Q(dw) \right] \right)
\end{equation}

\noindent where $Q$ is a L\'{e}vy measure on $[0,\infty)$ and satisfying $\int_{(0,\infty)}(1 \wedge x)Q(dx)<\infty$ (\cite{Bertoin_1997}, p.72). By the L\'{e}vy-Ito integral representation, we may express $W(t)$ directly as:

\begin{align}
    W(t) &= \int_{(0,\infty)} w N([0, t], dw) \label{eq:Levy_direct}
\end{align}

\noindent Here $N$ is a bivariate point process having mean measure $Lebesgue \times Q$ on $[0,T]\times (0,\infty)$, which can be conveniently expressed as:

\[
N=\sum_{i=1}^\infty \delta_{V_i,X_i}
\]

\noindent where  $\{V_i\in[0,T]\}$ are i.i.d. uniform random variables which give the times of arrival of jumps, $\{X_i\}$ are the size of the jumps and $\delta_{V,X}$ is the Dirac measure centered at time $V$ and jump size $X$.

Substituting $N$ directly into (\ref{eq:Levy_direct}) leads to

\begin{equation}
W(t)=\sum_{i=1}^\infty X_i{\cal I}_{V_i\leq t},\,\,\,as\label{W_realise}
\end{equation}

\noindent  The almost sure convergence of this series to $\{W(t)\}$ is proven in \cite{Rosinski_2001}. Most of the new material in this paper is concerned with generating the jump sizes $\{X_i\}$ for the GIG case. We will thus usually refer to the point process as just the set of jump sizes,  $N=\{X_i\}$, but of course corresponding jump times $\{V_i\}$ will need to be sampled as above in all cases in order to realise the process according to (\ref{W_realise}).

In principle it may be possible to simulate directly from the point process $N$, but in practice the potentially infinite number of jumps in any finite time interval make this impossible. If the jumps can be ordered by size, however, it may be possible to simulate all of the significant jumps and ignore or approximate the residual error from omitting the smallest jumps. It turns out that this can be done in a very convenient way, using the well known approach of \cite{Ferguson_Klass,Rosinski_2001,WolpertIckstadt1998}. The starting point is the simulation of the epochs of a unit rate Poisson process $\{\Gamma_i\}_{i\geq 1}$. The intensity function of this process is of course $\lambda_\Gamma=1$, and the resulting realisations contain almost surely an infinite number of terms. We know how to simulate an arbitrarily large number of ordered terms from this process by repeatedly generating standard exponential random variables and calculating the cumulative sum of these to obtain an ordered $\Gamma_i$ sequence. Now, define the upper tail probability of the L\'{e}vy measure as
\[
 Q^+(x)=Q ([x,\infty))
\]
and a corresponding non-increasing function $h(.)$ as the inverse tail probability:
\[
h(\gamma)={Q^+}^{-1}(\gamma)\,. 
\] 
Then, the following point process converges a.s. to $N$ \cite{Rosinski_2001}:
\[
    \sum_{i=1}^\infty \delta_{V_i,h(\Gamma_i)}
\]
and the corresponding convergent representation of the L\'{e}vy process is (neglecting the compensator term $c_i$, which is zero for all cases considered here):
\begin{equation}
W(t)=\sum_{i=1}^\infty h(\Gamma_i){\cal I}_{V_i\leq t},\,\, as\label{shot_noise_gen}
\end{equation}
 Thus, to generate points directly from $N$ by this method it is necessary to be able to compute $h(\gamma)={Q^+}^{-1}(\gamma)$ explicitly. In the cases considered here it will not be possible to do this and instead an indirect approach is adopted, using {\em thinning\/} or {\/rejection sampling}  
 \cite{Lewis_Shedler_1979,Rosinski_2001} of more tractable point processes for which $h(\gamma)$ is directly available. In particular, we seek a `bounding' process $N_0$ having L\'{e}vy measure $Q_0$ and satisfying $dQ_0(x)/dQ(x)\geq 1$ $\forall x\in(0,\infty)$; then, realisations of $N_0$ are thinned with probability $dQ(x)/dQ_0(x)$ in order to obtain samples from the desired process $N$. In all cases considered here the L\'{e}vy measure possesses a density function, which we also denote by $Q(x)$ (using the minor abuse of notation `$dQ(x)=Q(x)dx$') and the required bounding condition is then $Q_0(x)\geq Q(x)$, with associated thinning probability $Q(x)/Q_0(x)$.

Two bounding processes are used extensively in the methods of this paper, the tempered stable (TS) and gamma processes, which may be simulated by standard shot noise methods as follows:

\subsection{Tempered stable point process}
\label{TS_sample}
In the tempered stable (TS) case the L\'{e}vy density is, for $\alpha\in(0,1)$ \cite{Shephard_BN_2001} (see also \cite{Rosinski_2007})

\begin{equation}
    Q(x) = Cx^{-1-\alpha} e^{-\beta x}, \quad \quad \quad x > 0   \label{Q_temp_stable}
\end{equation}

\noindent Several possible approaches to simulation of sample paths from tempered stable processes were proposed in \cite{Rosinski_2001,Rosinski_2007} and compared in \cite{Imai2011OnFT}, which recommends the use of the inverse L\'{e}vy measure approach over thinning and rejection sampling methods. For the inverse L\'{e}vy measure approach the tail probability may be calculated in terms of gamma functions, but is not easily inverted, and numerical approximations are needed \cite{Imai2011OnFT}. We thus adopt a thinning approach \cite{Rosinski_2001} in which the L\'{e}vy density is factorised into a positive $\alpha$-stable process  with L\'{e}vy density $Q_0(x)=Cx^{-1-\alpha}$ \cite{Samorodnitsky_Taqqu_1994} and a tempering function $e^{-\beta x}$. The stable law process has tail mass $Q_0^+(x)=\frac{C}{\alpha}x^{-\alpha}$ and hence $h(\gamma)=Q_0^+(\gamma)^{-1}=\left(\frac{\alpha\gamma}{C}\right)^{-1/\alpha}$. Having simulated the stable point process with rate function $Q_0$, points are individually selected (thinned) with probability $e^{-\beta x}$, otherwise deleted.

The recipe for generating the tempered stable process $N_{TS}$ is then:
\begin{algorithm}
\caption{Generation of tempered stable process}
\label{gen_temp_stable}
\begin{enumerate}
    \item Assign $N_{TS}=\emptyset$.
    \item Generate the epochs of a unit rate Poisson process, $\{\Gamma_i;\,i=1,2,3...\}$
    \item For $i=1,2,3...$
    \begin{enumerate}
    \item Compute    $x_i=\left(\frac{\alpha\Gamma_i}{C}\right)^{-1/\alpha}$.
    \item
 With probability $e^{-\beta x_i}$, accept $x_i$ and assign $N_{TS}=N_{TS}\cup x_i$.
 \end{enumerate}
 \item For each jump $x_i\in N_{TS}$, generate independently a corresponding jump time $v_i\sim {\cal U}[0,T]$.
 \item Form a realisation of the tempered stable L\'{e}vy process as:
 \[
 w_{TS}(t)=\sum_{i=1}^\infty x_i{\cal I}_{v_i\leq t}
 \]
  \end{enumerate}
\end{algorithm}

In practice $i$ is truncated at some large value and an approximate realisation results.

\subsection{Gamma process}
\label{Gamma_process}
The  L\'{e}vy density for the Gamma process is:
\[
Q(x)={C}{x^{-1}}e^{-\beta x}, \quad \quad \quad x > 0
\]
\cite{Rosinski_2001} suggests four possible sampling schemes for this process. The tail probability is the Exponential integral, which would require numerical inversion, see \cite{Wolpert_Ickstadt_1998}. Instead we adopt the thinning (also called the `rejection method' in \cite{Rosinski_2001}) version of this in which a dominating point process is chosen as $Q_0(x)=\frac{C}{x}(1+\beta x)^{-1}$. The tail probability for this is $Q_0^+(x)=C\log\left(\beta^{-1} x^{-1}+1\right)$ and hence $h(\gamma) = \frac{1}{\beta \left( \exp(\gamma / C) - 1 \right)}$. Points are then  thinned with probability $(1+\beta x) \exp(-\beta x) \leq 1$. As reported in \cite{Rosinski_2001}, this thinning method is highly effective, with very few point rejections observed. The recipe for generating the gamma process $N_{Ga}$ is then given in Algorithm \ref{gen_gamma}.

\begin{algorithm}
\caption{Generation of gamma process}
\label{gen_gamma}
\begin{enumerate}
    \item Assign $N_{Ga}=\emptyset$.
    \item Generate the epochs of a unit rate Poisson process, $\{\Gamma_i;\,i=1,2,3...\}$
    \item For $i=1,2,3...$
    \begin{enumerate}
    \item Compute    $x_i=\frac{1}{\beta \left( \exp(\Gamma_i / C) - 1 \right)}$.
    \item
 With probability $(1+\beta x) \exp(-\beta x_i)$, accept $x_i$ and assign $N_{Ga}=N_{Ga}\cup x_i$.
 \end{enumerate}
 \item For each jump $x_i\in N_{Ga}$, generate independently a corresponding jump time $v_i\sim {\cal U}[0,T]$.
 \item Form a realisation of the Gamma L\'{e}vy process as:
 \[
 w_{Ga}(t)=\sum_{i=1}^\infty x_i{\cal I}_{v_i\leq t}
 \]
  \end{enumerate}
 \end{algorithm}

As before $i$ is truncated at some large value, yielding an approximate realisation of the gamma process.

\section{Simulation from the generalised inverse Gaussian L\'{e}vy process}

In this section the GIG L\'{e}vy density is presented and a general scheme is presented that will enable simulation from the GIG process. 

The L\'{e}vy density for the generalised inverse Gaussian (GIG) process is given by \cite{Eberlein2003}:

\begin{equation}
\frac{e^{-x\gamma^2/2}}{x}\left[\int_0^\infty\frac{e^{-xy}}{\pi^2y|H_{|\lambda|}(\delta\sqrt{2y})|^2}dy+\text{max}(0,\lambda)\right],\,\,x>0
\label{GIG_levy_density}
\end{equation}
where
$H_{\lambda}(z)=J_{\lambda}(z)+iY_{\lambda}(z)$ is the  Hankel function of the first kind ($z$ is always real-valued in the current context), $J_{\lambda}(z)$ is the Bessel function of the first kind, and  $Y_{\lambda}(z)$ is the  Bessel function of the second kind.

The GIG L\'{e}vy density comprises two terms: the initial integral, which we denote $Q_{GIG}(x)$:
\[
Q_{GIG}(x)=\frac{e^{-x\gamma^2/2}}{x}\int_0^\infty\frac{e^{-xy}}{\pi^2y|H_{|\lambda|}(\delta\sqrt{2y})|^2}dy,\,\,x>0
\]
added to a second term, present only for $\lambda>0$:
\begin{equation}
    \frac{e^{-x\gamma^2/2}}{x}\text{max}(0,\lambda),\,\,x>0 \label{gamma_process_positive_lambda}
\end{equation}
This second term is a Gamma process that may be straightforwardly simulated using the standard methods of \ref{Gamma_process} and added to the simulation of points from the first term $Q_{GIG}(x)$. Hence we will neglect this second term for now. It will be convenient to rewrite $Q_{GIG}(x)$ using the substitution $z=\delta\sqrt{2y}$ as:

\begin{align*}
    Q_{GIG}(x)&=\frac{e^{-x\gamma^2/2}}{x}\int_0^\infty\frac{e^{-xy}}{\pi^2y|H_{|\lambda|}(\delta\sqrt{2y})|^2}dy \\
    &=\frac{2 e^{-x\gamma^2/2}}{\pi^2x}\int_0^\infty\frac{e^{-\frac{z^2x}{2\delta^2}}}{z|H_{|\lambda|}(z)|^2}dz
\end{align*}

Note that this integral is known elsewhere as the Jaeger integral, which finds application in diffusive transport \cite{Freitas2018}. Beyond our direct interest in GIG processes, there is significant interest in approximating these integrals accurately, and our bounding approach is likely to provide accurate bounds and approximations which may be compared with those proposed in \cite{Freitas2018}.

Throughout this paper we will consider only positive and real-valued variables $x$, $y$ and $z$, as the complex versions of these functions are not required in the present context.

Our general scheme is to consider the following intensity function associated with a bivariate point process on $(0,\infty)\times (0,\infty)$:
\begin{equation}
Q_{GIG}(x,z)=\frac{2 e^{-x\gamma^2/2}}{\pi^2x}\frac{e^{-\frac{z^2x}{2\delta^2}}}{z|H_{|\lambda|}(z)|^2}\label{Q_bivariate}
\end{equation}
which has by construction the GIG process as its marginal:
\[
Q_{GIG}(x)=\int_{0}^\infty Q_{GIG}(x,z)dz
\]
We propose to simulate points directly from this bivariate process, hence avoiding any direct evaluation of the Jaeger integral. Since $Q_{GIG}(x,z)$ is intractable itself for simulation, dominating processes $Q^0_{GIG}(x,z)\geq Q_{GIG}(x,z)$ are constructed and points sampled from $Q^0_{GIG}(x,z)$ are thinned with probability\newline $Q_{GIG}(x,z) / Q^0_{GIG}(x,z)$. Thus a significant part of our approach is in constructing suitable dominating functions that are tractable for simulation, and this is achieved by studying the properties of the Jaeger integral. 

The first set of bounds are obtained from the basic properties of the Hankel function \cite{Watson1944} and will lead in particular to a simulation algorithm for the case $|\lambda|\geq 0.5$. Properties of the modulus of the Hankel function are mainly derived from the Nicholson integral representation \cite{Watson1944},
\[
 |H_\nu(z)|^2=\frac{8}{\pi^2}\int_0^\infty K_0(2z\sinh t)\cosh (2\nu t) dt 
\]
\noindent where $K_0$ is the modified Bessel function of the second kind. In particular this leads to an asymptotic ($z\rightarrow \infty$) series expansion (\cite{Watson1944}, Section 13.75):
\begin{align*}
    |H_{\nu}(z)|^{2} \sim &\frac{2}{\pi z} \bigg( 1+\frac{4\nu^{2}-1}{2(2z)^{2}} + \frac{1\cdot 3(4\nu^{2}-1)(4\nu^{2}-9)}{2\cdot 4(2z)^{4}} \\
    &+  \frac{1\cdot 3\cdot 5(4\nu^{2}-1)(4\nu^{2}-9)(4\nu^{2}-25)}{2\cdot 4\cdot 6(2z)^{6}}\,\,... \bigg) \,,
\end{align*}

\noindent from which the well known asymptotic value is obtained,
\[
\underset{z\rightarrow \infty}{\text{lim}} z|H_{\nu}(z)|^{2}=\frac{2}{\pi}
\]
From the Nicholson integral the following important property can also be derived ( \cite{Watson1944}, Section 13.74):

\begin{property}
\label{Property_monotonicity}
For any real, positive $z$, $z|H_{\nu}(z)|^{2}$ is a {\em decreasing\/} function of $z$ for $\nu>0.5$ and an {\em increasing\/} function of $z$ when $0 <\nu<0.5$.

\noindent For the case $\nu=0.5$,  $z|H_{\nu}(z)|^{2}=2/\pi$, i.e. a constant.
\end{property}
These basic properties are used to prove the following bounds:
\begin{theorem}
\label{Lemma_1}
For any positive $z$ and fixed $|\lambda|\geq 0.5$, the following bound applies:
\begin{equation}
Q_{GIG}(x,z)\leq \frac{ e^{-x\gamma^2/2}}{\pi x}{e^{-\frac{z^2x}{2\delta^2}}}\label{Q_GIG_bound}
\end{equation}
and, for $|\lambda|\leq 0.5$,
\begin{equation}
Q_{GIG}(x,z)\geq\frac{ e^{-x\gamma^2/2}}{\pi x}{e^{-\frac{z^2x}{2\delta^2}}}\label{Q_GIG_bound_l_0_5}
\end{equation}
with equality holding in both cases when $|\lambda|=0.5$.
\end{theorem}
\begin{proof}
Property \ref{Property_monotonicity} and the limiting value of $2/\pi$ lead to
\begin{align}
z|H_\nu(z)|^2\begin{cases}\geq \frac{2}{\pi},& \nu\geq 0.5\label{H_bound}\\\leq\frac{2}{\pi },&\nu\leq 0.5\end{cases}
\end{align}
    The results follow by substitution of (\ref{H_bound}) into $Q_{GIG}(x,z)$ (\ref{Q_bivariate}).
\end{proof}

\begin{corollary}
\label{cor_GIG_bound}
 The bound in (\ref{Q_GIG_bound}), applicable for $|\lambda|\geq 0.5$, can be rewritten as
 \begin{equation}
 \frac{ e^{-x\gamma^2/2}}{\pi x}{e^{-\frac{z^2x}{2\delta^2}}}=\frac{ {\delta\Gamma(1/2)}e^{-x\gamma^2/2}}{\sqrt{2}\pi x^{3/2}}\sqrt{{Ga}}\left(z|1/2,\frac{x}{2\delta^2}\right)\label{temp_stable}
 \end{equation}
 where $\sqrt{Ga}$ is the square-root gamma density, i.e. the density of $X^{0.5}$ when $X\sim Ga(x|\alpha,\beta)$, having probability density function
 \[
        \sqrt{\text{Ga}}(x|\alpha, \beta) = \frac{2 \beta^{\alpha}}{\Gamma(\alpha)} x^{2\alpha-1} e^{-\beta x^2} 
 \]
 
 \end{corollary}
 
\begin{remark}
It can be seen immediately that (\ref{temp_stable}) corresponds ${}$ marginally to a tempered stable process in $x$, and conditionally to a $\sqrt{Ga}$ density for $z$, a tractable feature that will enable sampling from the dominating bivariate point process. In fact, this decomposition and that of Corollory \ref{cor_piecewise} are the key point for our new GIG simulation methods. We are here decomposing the bivariate point process in $(x,z)$ as a {\em marked} point process comprising a marginal Poisson process for $x$ and a conditional mark random variable $z$ (\cite{Cont:2003}, Section 2.6.4). The important point here is that the conditional distributions of $z|x$ are integrable and tractable probability density functions and hence a marginal-conditional sampling procedure may be adopted constructively to simulate from the joint point process in $(x,z)$.
\end{remark}

\begin{remark}
Integration\/  with respect to $z$ leads to a simple upper bound on the GIG L\'{e}vy density:
\begin{equation}
\frac{ {\delta\Gamma(1/2)}e^{-x\gamma^2/2}}{\sqrt{2}\pi x^{3/2}} \label{GIG_upper}
\end{equation}
which was also obtained by \cite{Zhang_2011}, and used in a rejection sampling procedure that requires a direct evaluation of the Jaeger integral, in contrast with our approach. For $|\lambda|\leq 0.5$, a similar argument leads to a lower bound with the same formula as (\ref{GIG_upper}).
 \end{remark}

\begin{remark}
\noindent The first bound for $\nu\geq 0.5$ in (\ref{H_bound}) will be used shortly as an 
envelope function for rejection sampling in the case $|\lambda|\geq 0.5$ using Corollary \ref{cor_GIG_bound}. The second bound in (\ref{H_bound}) for $\nu\leq 0.5$ will be adapted in Theorem \ref{Lemma_2} to develop more sophisticated bounds in this parameter range.
\end{remark}
A second set of bounds can be stated as follows:
\begin{theorem}
\label{Lemma_2}
Choose a point $z_0\in(0,\infty)$ and compute $H_0=z_0|H_{\nu}(z_0)|^2$. This will define the corner point on a  piecewise lower or upper bound. Define now $z_1 = \left(\frac{ 2^{1-2\nu}\pi}{\Gamma^2(\nu)}\right)^{1/(1-2\nu)}$ and define the following functions:

\[
A(z)=\begin{cases}\frac{2}{\pi}\left(\frac{z_1}{z}\right)^{2\nu-1},&z < z_1 \\ \frac{2}{\pi},&z\geq z_1 \end{cases}
\]
and
\[
B(z)=\begin{cases}H_0\left(\frac{z_0}{z}\right)^{2\nu-1},&z<z_0\\H_0,&z\geq z_0\end{cases}
\]
Then, for $0<\nu\leq 0.5$,
\begin{equation}
A(z)\geq z|H_{\nu}(z)|^2\geq B(z) \label{bound_lam_leq_0_5}
\end{equation}
and for $\nu\geq 0.5$,
\begin{equation}
A(z)\leq z|H_{\nu}(z)|^2\leq B(z) \label{bound_lam_geq_0_5}
\end{equation}
with all inequalities becoming equalities when $\nu=0.5$, and both $A(z)$ bounds (left side inequalities) becoming tight at $z=0$ and $z=\infty$. 
\end{theorem}
\begin{proof}
 First note that $z|H_{\nu}(z)|^2$ is an increasing function for $0 < \nu < 0.5$ and decreasing for $\nu > 0.5$ as stated in Property \ref{Property_monotonicity}. When combined with the tight bound of $2/\pi$ as $z\rightarrow \infty$ (\ref{H_bound}) this leads directly to the $2/\pi$ components of $A(z)$.  
 Similarly the first part of the bound $A(z)= z(\Gamma(\nu)/\pi)^2\left(\frac{2}{z}\right)^{2\nu}$, is justified since \cite{Freitas2018} proves that
 $z^{2\nu}|H_{\nu}(z)|^2$ is increasing when $\nu>0.5$ and decreasing when $\nu<0.5$, with a tight bound of $(\Gamma(\nu)/\pi)^2 2^{2\nu}$ as $z\rightarrow 0$.
 Thus the left hand side of the inequalities, $A(z)$, are established. 
 
 The right hand side of the inequalities, $B(z)$, are proven by the monotonicity and sign of gradient for both $z^{2\nu}|H_{\nu}(z)|^2$ and $z|H_{\nu}(z)|^2$. We may choose an arbitrary corner point $(z_0,H_0)$ on  
 $z|H_{\nu}(z)|^2$ and monotonicity of $z|H_{\nu}(z)|^2$ (Property \ref{Property_monotonicity}) immediately implies that, for $z\geq z_0$, \[z|H_{\nu}(z)|^2\begin{cases}\geq H_0,&0<\nu\leq 0.5\\ \leq H_0,&\nu\geq 0.5\end{cases}\]
 Moreover, monotonicity and sign of gradient of $z^{2\nu}|H_{\nu}(z)|^2$ \cite{Freitas2018} imply that

\begin{figure*}
  \includegraphics[width=\textwidth]{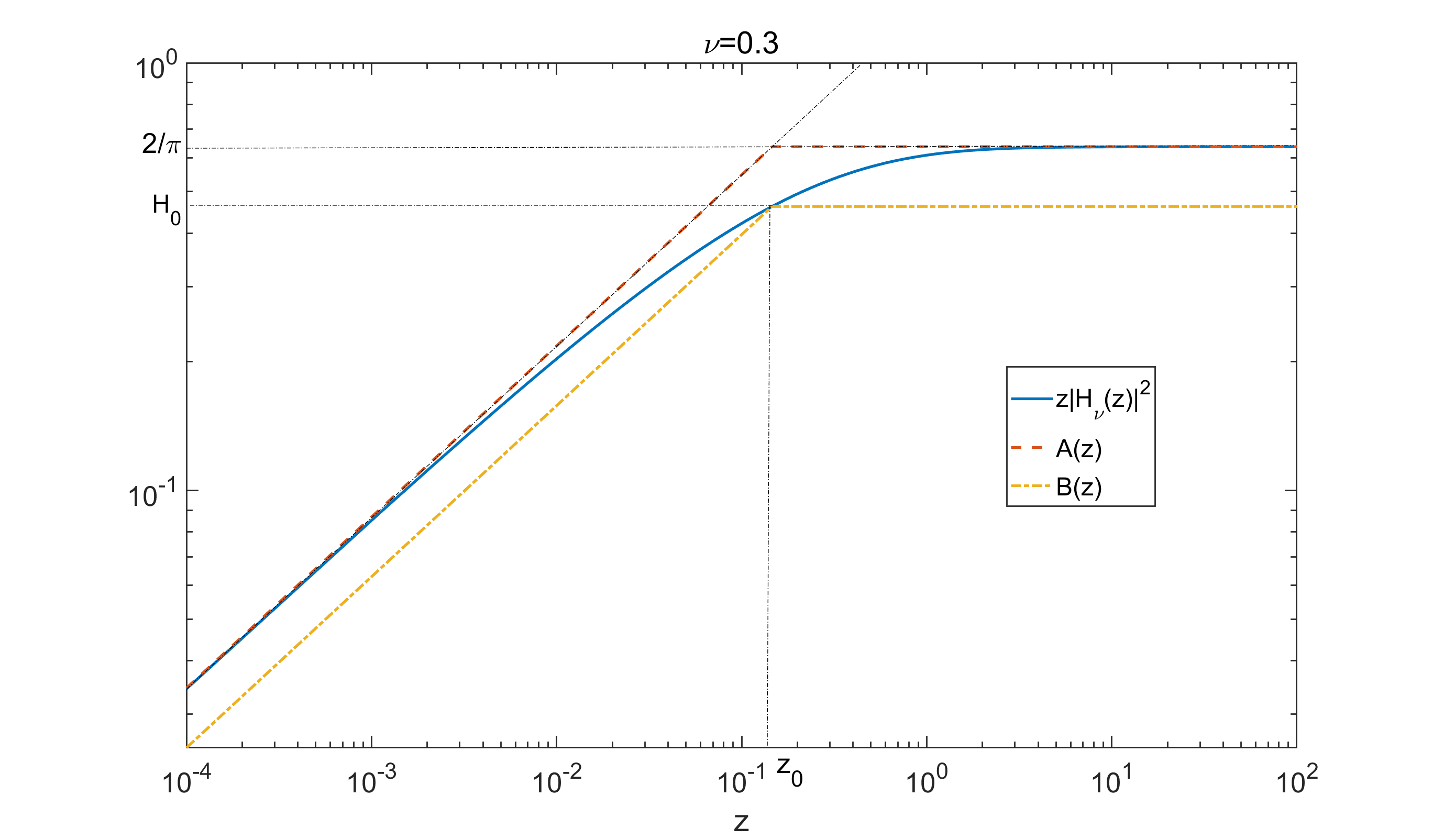}
\caption{Plot of Bessel function bounds, $\nu=0.3$. $z_0$ set equal to $z_1$.}
\label{bounds_lambda_0_3}       
\end{figure*}

\begin{figure*}
  \includegraphics[width=\textwidth]{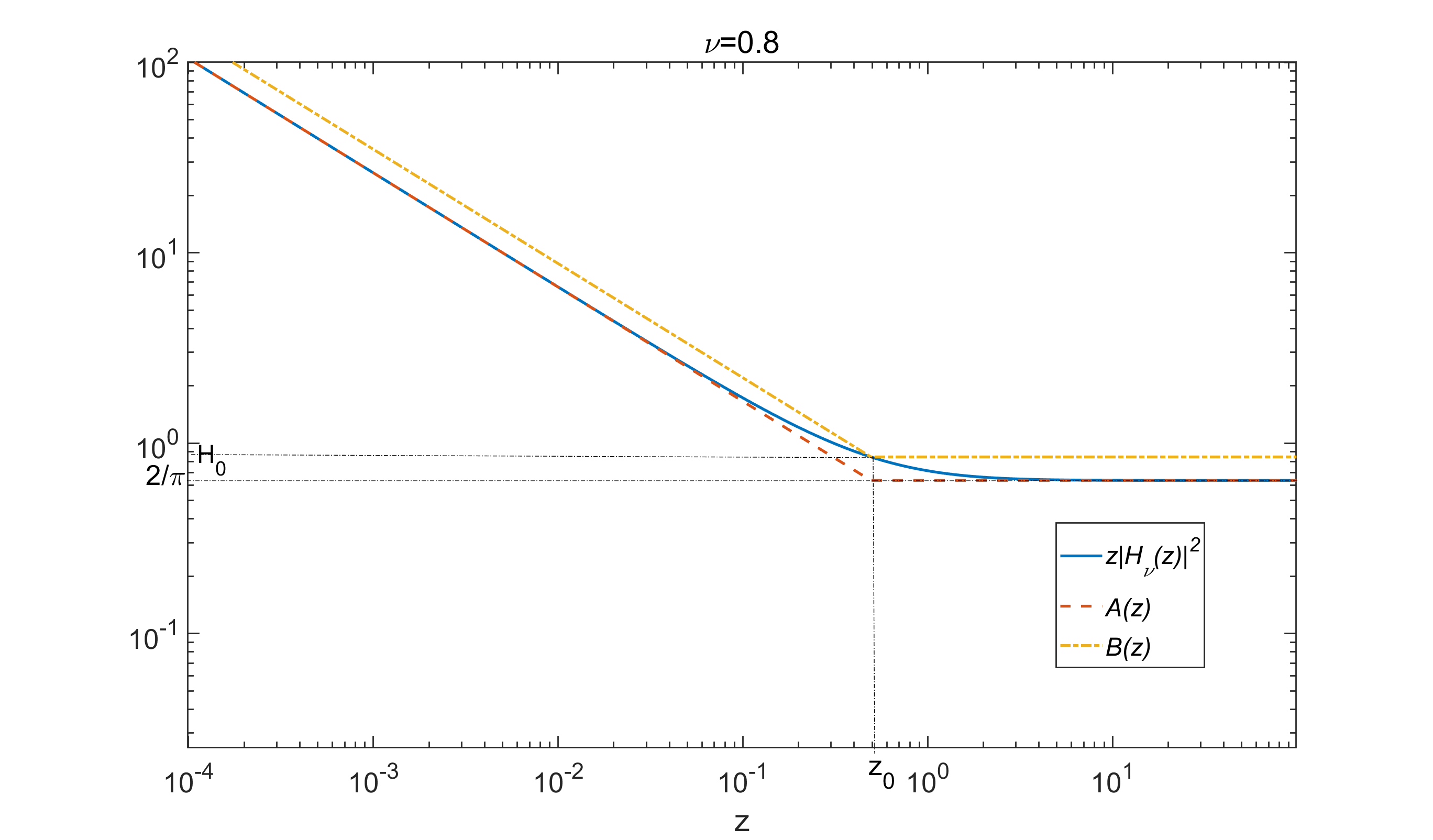}
\caption{Plot of Bessel function bounds, $\nu=0.8$. $z_0$ set equal to $z_1$.}
\label{bounds_lambda_0_8}       
\end{figure*}

 \[
 z^{2\nu}|H_{\nu}(z)|^2\begin{cases}\geq z^{2\nu-1}H_0,&0<\nu\leq 0.5\\ \leq z^{2\nu-1} H_0,&\nu\geq 0.5\end{cases}
 \]
 from which the bounds $B(z)$ are immediately obtained.
 \end{proof}
 \begin{remark}
 The proven bounds can be clearly visualised on a log-log plot, which highlights the asymptotic behaviour of the functions and bounds as $z\rightarrow 0 $ and $z\rightarrow \infty$, see Figs. \ref{bounds_lambda_0_3} and \ref{bounds_lambda_0_8}.
 \end{remark}
 \begin{remark}
 The choice of $z_0\in (0,\infty)$ is arbitrary. However, its value will impact the tightness of the bounding functions $B(z)$  and will hence impact the effectiveness of our subsequent sampling algorithms and integral approximations. A suitable generic choice was  found to be at the same $z$ value as the corner point of the corresponding $A(z)$  bounds, i.e. set $z_0=z_1$, as plotted in Figs. \ref{bounds_lambda_0_3} and \ref{bounds_lambda_0_8}, though further optimisation may be possible in applications.
 \end{remark}
 
\begin{figure*}[ht]
\centering
\includegraphics[width=0.9\textwidth]{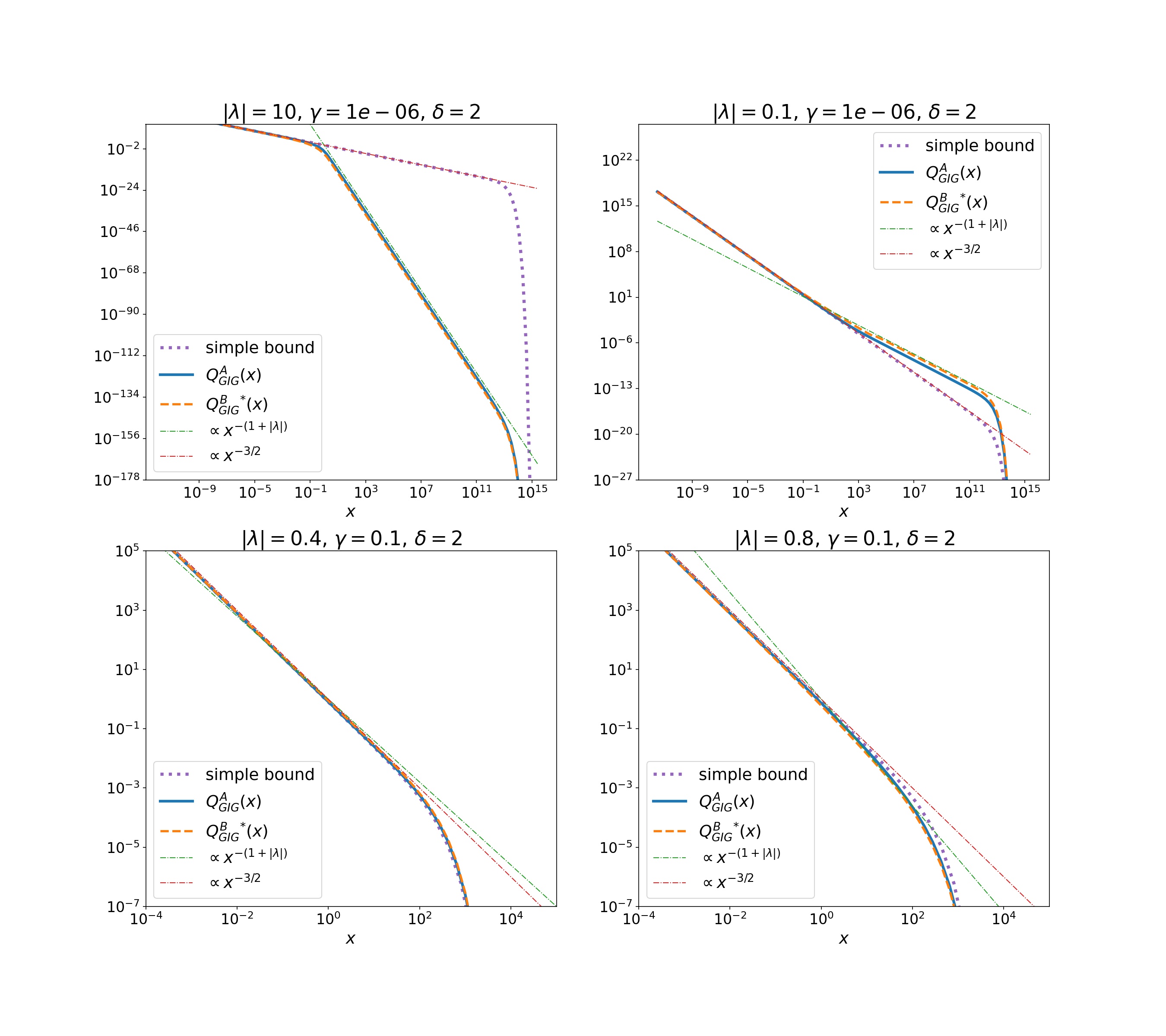}
\caption{Plot of optimised upper and lower bounds on $Q_{GIG}(x)$, for various parameter settings. We use the following shorthand for the optimised $Q^B$ bounds: ${Q^B_{GIG}}^{*}(x) = \underset{z_0\in (0,\infty)}{\mathrm{max/min}}\left\{Q^B_{GIG}(x)\right\}$, where $\text{min}$ applies for $|\lambda|<0.5$ and $\text{max}$ applies for $|\lambda|\geq 0.5$. Overlaid also are the simple approximation (\ref{H_bound}) and the trends $x^{-3/2}$ and $x^{-(1+|\lambda|)}$.}
 \label{jaeger_integral}
\end{figure*}

\begin{corollary}
\label{cor_piecewise}
For the case $|\lambda|<0.5$, the right hand bound in (\ref{bound_lam_leq_0_5}), $B(z)$,  can be substituted into $Q_{GIG}(x,z)$ (\ref{Q_bivariate}) and rearranged to obtain:
\begin{align}
Q_{GIG}(x,z) &=
\frac{2 e^{-x\gamma^2/2}}{\pi^2x}\frac{e^{-\frac{z^2x}{2\delta^2}}}{z|H_{|\lambda|}(z)|^2} \\ 
&\leq \frac{ 2 e^{-x\gamma^2/2}}{\pi^2x}
\begin{cases}
\frac{z^{2|\lambda|-1}e^{-\frac{z^2x}{2\delta^2}}}{H_0z_0^{2|\lambda|-1}},&0<z<z_0\\
\frac{e^{-\frac{z^2x}{\delta^2}}}{H_0},&\text{otherwise}
\end{cases}\nonumber
\end{align}

\noindent Furthermore, the two piecewise sections of this bound may be factorised in terms of left- and right-truncated square-root gamma densities:
\begin{align}
    Q_{GIG}(x,z) &\leq \frac{ e^{-x\gamma^2/2}}{\pi^2x^{1+|\lambda|}}\frac{(2\delta^2)^{|\lambda|}\gamma(|\lambda|,z_0^2x/(2\delta^2))}{H_0z_0^{2|\lambda|-1}} \frac{\Gamma(|\lambda|)\sqrt{\text{Ga}}
(z||\lambda|,x/(2\delta^2))}{\gamma(|\lambda|,z_0^2x/(2\delta^2))}{\cal I}_{0<z<z_0} \nonumber
\\&+\frac{ e^{-x\gamma^2/2}}{\pi^2x^{3/2}}\frac{(2\delta^2)^{0.5}\Gamma(0.5,z_0^2x/(2\delta^2))}{H_0} \frac{\Gamma(0.5)\sqrt{\text{Ga}}
(z|0.5,x/(2\delta^2))}{\Gamma(0.5,z_0^2x/(2\delta^2))} {\cal I}_{z\geq z_0} \label{GIG_bound_lam_le_0_5}
\end{align}
where the truncated square-root gamma densities are, with their associated normalising constants:
\[
\frac{\Gamma(|\lambda|)\sqrt{\text{Ga}}
(z||\lambda|,x/(2\delta^2))}{\gamma(|\lambda|,z_0^2x/(2\delta^2))}{\cal I}_{0<z<z_0}\,,
\]
\[
\frac{\Gamma(0.5)\sqrt{\text{Ga}}
(z|0.5,x/(2\delta^2))}{\Gamma(0.5,z_0^2x/(2\delta^2))} {\cal I}_{z\geq z_0}
\]
and lower/upper incomplete gamma functions are defined in the usual way, for $\text{Re}(s) > 0$, as:
\begin{align*}
    \gamma(s,x) = \int_{0}^{x} t^{s-1} e^{-t} dt,\,\,\,\,
    \Gamma(s,x) = \int_{x}^{\infty} t^{s-1} e^{-t} dt
\end{align*}

\noindent $Q_{GIG}(x,z)$ can thus be split up into two tractable point processes for simulation: a first, $N_1$, comprising a modified tempered $|\lambda|$-stable process with truncated $\sqrt{Ga}$ conditional for $z$; and a second, $N_2$, comprising a modified tempered 0.5-stable process with a further truncated $\sqrt{Ga}$ conditional for $z$. We will later use this union of point processes as the dominating L\'{e}vy measure in simulation of the $|\lambda|<0.5$ case.

 \end{corollary}
  \begin{corollary}
 Integration of the bounding function (\ref{GIG_bound_lam_le_0_5}) with respect to $z$ allows a more sophisticated estimate for the Jaeger integral and therefore the L\'{e}vy density for the GIG process, which is an upper bound for $|\lambda|<0.5$ and a lower bound for $|\lambda|>0.5$, following the direction of the right hand inequalities in (\ref{GIG_bound_lam_le_0_5}) and (\ref{bound_lam_geq_0_5}). A similar procedure inserting the left hand ($A(z)$) inequalities from (\ref{GIG_bound_lam_le_0_5}) and (\ref{bound_lam_geq_0_5}) yields the corresponding lower ($|\lambda|<0.5$) and upper ($|\lambda|>0.5$) bounds. Define first the  bounding functions $Q^A_{GIG}(x)$ and $Q^B_{GIG}(x)$, corresponding to the bounds $A(z)$ and $B(z)$ as follows:

\begin{align*}
    Q^A_{GIG}(x) = 
\frac{ e^{-x\gamma^2/2}}{\pi x^{1+|\lambda|}}\frac{(2\delta^2)^{|\lambda|}\gamma(|\lambda|,z_1^2x/(2\delta^2))}{2z_1^{2|\lambda|-1}}+\frac{ e^{-x\gamma^2/2}}{\pi x^{3/2}}\frac{(2\delta^2)^{0.5}\Gamma(0.5,z_1^2x/(2\delta^2))}{2}
\end{align*}

\begin{align*}
    Q^B_{GIG}(x) = 
\frac{ e^{-x\gamma^2/2}}{\pi^2 x^{1+|\lambda|}}\frac{(2\delta^2)^{|\lambda|}\gamma(|\lambda|,z_0^2x/(2\delta^2))}{H_0z_0^{2|\lambda|-1}}+\frac{ e^{-x\gamma^2/2}}{\pi^2x^{3/2}}\frac{(2\delta^2)^{0.5}\Gamma(0.5,z_0^2x/(2\delta^2))}{H_0}
\end{align*}
These are obtained by substituting the bounds $A(z)$ or $B(z)$ into the expression for $Q_{GIG}(x,z)$ (\ref{Q_bivariate}) and integrating with respect to $z$.
 
 Noting that $z_0\in (0,\infty)$ can be chosen arbitrarily, the $Q^B_{GIG}$ estimates may be improved by optimising with respect to $z_0$ to either maximise ($|\lambda|< 0.5$) or minimise ($|\lambda|> 0.5$) $Q^B_{GIG}(x)$ at each point $x$. 
 
 Then, following the direction of the  inequalities in (\ref{bound_lam_leq_0_5}) and (\ref{bound_lam_geq_0_5})
 we obtain the following estimates of $Q_{GIG}(x)$:
 \[
 Q^A_{GIG}(x)\leq Q_{GIG}(x)\leq \underset{z_0\in (0,\infty)}{\mathrm{min}}\left\{Q^B_{GIG}(x)\right\},\,\,\,\,|\lambda|\leq 0.5
 \]
\[
 Q^A_{GIG}(x)\geq Q_{GIG}(x)\geq \underset{z_0\in (0,\infty)}{\mathrm{max}}\left\{Q^B_{GIG}(x)\right\},\,\,\,\,|\lambda|\geq 0.5
 \]
 with all inequalities becoming equalities at $|\lambda|=0.5$. Optimisation to achieve the required maximum and minimum functions can be achieved by numerical search. 
 
 Example plots are given in Fig. \ref{jaeger_integral} in which the optimised lower and upper bounds are plotted, showing very close agreement (often indistinguishable by eye) in their estimation of $Q_{GIG}(x)$, over various parameter ranges and large $x$ ranges (i.e. the bounds are visually quite tight).  Overlaid are the $x^{-3/2}$  and $x^{-(1+|\lambda|)}$ trends and also the simple approximation of (\ref{GIG_upper}). 
  This simple approximation diverges significantly from our proposed bounds, particularly when $|\lambda|$ is not close to 0.5 (when $|\lambda|=0.5$ all bounds are equal to the true function $Q_{GIG}(x)$ for the inverse Gaussian (IG) process).
 
 




\end{corollary}

\subsection{Simulation of GIG processes with $|\lambda|\geq 0.5$}

In this section the specific algorithm applied for simulation in the case $|\lambda|\geq 0.5$ is detailed. In this parameter range it has been found very effective to use Theorem \ref{Lemma_1} and Corollary \ref{cor_GIG_bound}, so that the dominating bivariate process is:
\[
 Q^0_{GIG}(x,z)=\frac{ e^{-x\gamma^2/2}}{\pi x}{e^{-{(z^2x)}/{(2\delta^2)}}}
\]
\noindent which is simulated as a marked point process having factorised (marginal-conditional) intensity function (Corollary \ref{cor_GIG_bound}) as follows:
\[
Q^0_{GIG}(x,z)=\frac{ {\delta\Gamma(1/2)}e^{-x\gamma^2/2}}{\sqrt{2}\pi x^{3/2}}\sqrt{{Ga}}\left(z|1/2,\frac{x}{2\delta^2}\right)\label{temp_stable}
\]

The thinning probability for points drawn from the dominating point process is then:
\[
\frac{Q_{GIG}(x,z)}{Q^0_{GIG}(x,z)}=\frac{2}{\pi z|H_{|\lambda|}(z)|^2}\]

The procedure for generation of points from the process with intensity function $Q_{GIG}(x)$ is given in Algorithm \ref{Q_GIG_gen}. In the case of $\lambda > 0$, points generated from the process with intensity function shown in Eq. (\ref{gamma_process_positive_lambda}) are added to the set of points coming from $Q_{GIG}(x)$ to obtain jump sizes from the GIG process.

%
\begin{algorithm}
\caption{Simulation for $Q_{GIG}(x)$ when $|\lambda|\geq 0.5$}
\label{Q_GIG_gen}
\begin{enumerate}
    \item $N=\emptyset$

    \item Generate a large number of points from the tempered stable process whose intensity function is (first factor in  (\ref{temp_stable})): \[
    \frac{e^{-x\gamma^2/2}}{x^{3/2}}\frac{\delta\Gamma(1/2)}{\sqrt{2}\pi}\,,
    \]
    using Algorithm \ref{gen_temp_stable}, setting $C=\frac{\delta\Gamma(1/2)}{\sqrt{2}\pi}$, $\alpha=0.5$ and $\beta=\gamma^2/2$.
    
    \item For each point $x$, draw a random variate $z\sim \sqrt{\text{Ga}}(z|1/2,x/(2\delta^2))$,
    \item For each $(x,z)$, accept with probability 
    \[
   \frac{2}{\pi z|H_{|\lambda|}(z)|^2},
    \]
    i.e. set $N=N\cup x$.
\end{enumerate}    
\end{algorithm}
The procedure is completed as before by independently drawing an associated jump time $v\sim {\cal U}[0,T]$ for each accepted jump size (`point') $x\in N$. The value of the GIG L{\'e}vy process at $t$ is given in Eq. (\ref{W_realise}).


\subsubsection{Acceptance rate}

\begin{figure*}[ht]
\centering
\includegraphics[width=0.9\textwidth]{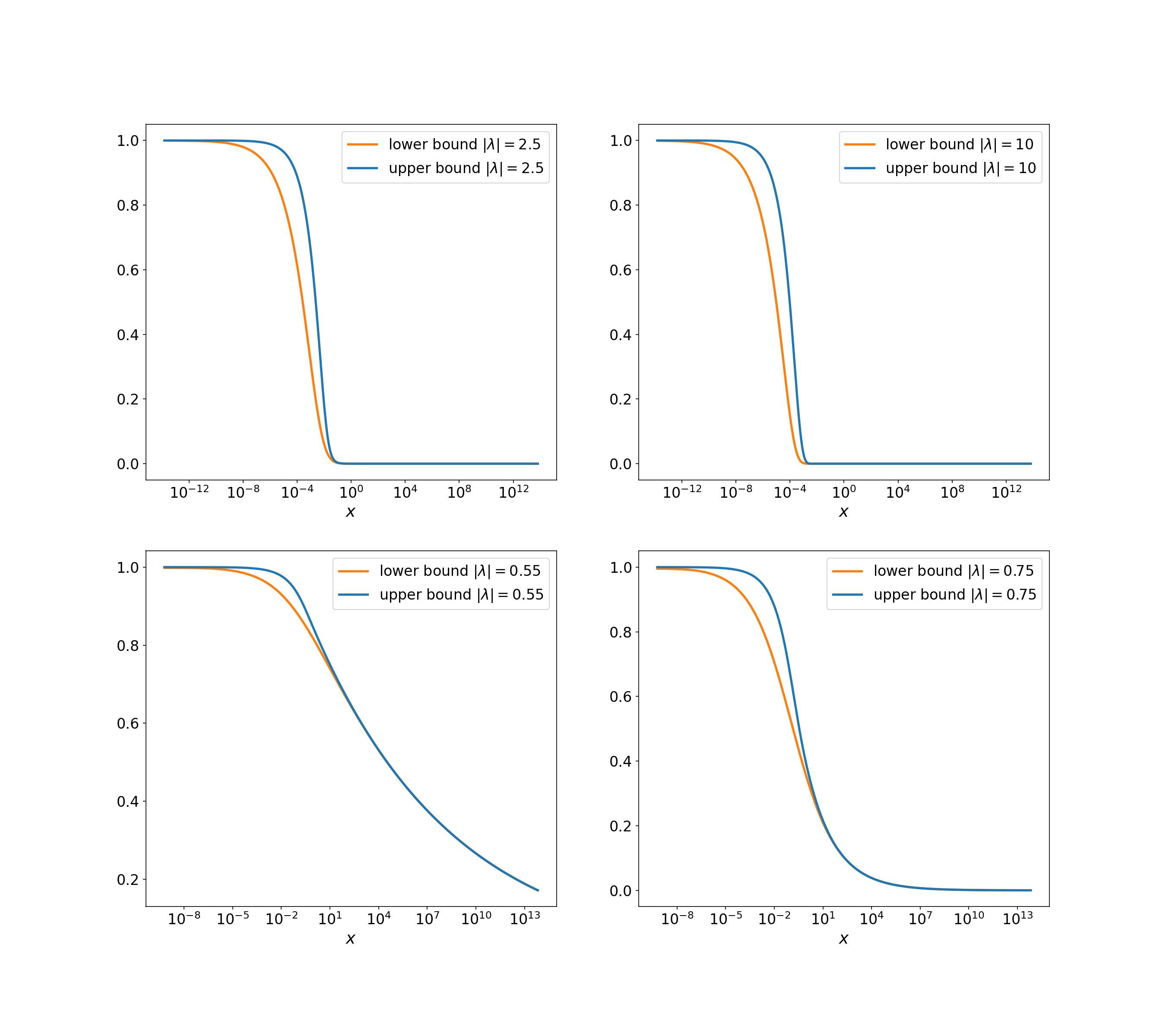}
\caption{Plot of upper and optimised lower bounds on $\rho(x)$, for various $|\lambda|>0.5$, showing that large jumps are rejected while with increasing probability small jumps are accepted. $\delta=0.1$ in all cases and bounds do not depend on $\gamma$}
 \label{rho_upper_lower_bounds}
\end{figure*}

The mean acceptance probability for fixed $x$ is:

\begin{align*}
\rho(x) &=E \left[ \frac{2}{\pi z|H_{|\lambda|}(z)|^2} \right] \\ &=\int_{0}^{\infty} \sqrt{\text{Ga}}(z|1/2,x/(2\delta^2))\frac{2}{\pi z|H_{|\lambda|}(z)|^2}dz
\end{align*}

\noindent which may be bounded using Theorem \ref{Lemma_2}. For the current parameter setting ($|\lambda|>0.5$) we have $A(z) \leq z |H_{|\lambda|}(z)|^2 \leq B(z)$. Lower and upper bounds are obtained by substituting $A(z)$ and $B(z)$ and then by direct integration to give:


\begin{equation}
\label{eqn:rho_lower_bound}
\begin{aligned}
\frac{2}{\pi H_0}& \left( \left(\frac{z_0^2x}{2\delta^2}\right)^{0.5-|\lambda|} \frac{\gamma(|\lambda|, \frac{z_0^2x}{2\delta^2})}{\Gamma(0.5)} + \frac{\Gamma(0.5, \frac{z_0^2x}{2\delta^2})}{\Gamma(0.5)} \right)
\leq \rho(x)\leq  \left(\frac{z_1^2x}{2\delta^2}\right)^{0.5-|\lambda|} \frac{\gamma(|\lambda|, \frac{z_1^2x}{2\delta^2})}{\Gamma(0.5)} + \frac{\Gamma(0.5, \frac{z_1^2x}{2\delta^2})}{\Gamma(0.5)}
\end{aligned}
\end{equation}
As noted before, the corner point $z_0\in (0,\infty)$ may be chosen arbitrarily, while $z_1$ is fixed. Hence the lower bound may be optimised with respect to $z_0$ at each $x$ value to obtain a tighter lower bound:
\begin{align*}
    \underset{z_0\in (0,\infty)}{\text{max}}\bigg\{ &\frac{2}{\pi H_0} \bigg( \bigg(\frac{z_0^2x}{2\delta^2}\bigg)^{0.5-|\lambda|} \frac{\gamma(|\lambda|, z_0^2x/(2\delta^2))}{\Gamma(0.5)} + \frac{\Gamma(0.5, z_0^2x/(2\delta^2))}{\Gamma(0.5)} \bigg) \bigg\} \leq \rho(x)
\end{align*}




The (numerically optimised) lower bound and (fixed) upper bound on the mean acceptance rate $\rho(x)$ are visualised  for different parameter settings in Fig. \ref{rho_upper_lower_bounds}. It can be observed, as expected from the inequalities (\ref{eqn:rho_lower_bound}), that the bounds are tight as $x \to 0$ and $x \to \infty$, and that acceptance rates get lower as $|\lambda|$ increases. In all cases though they indicate that only large jumps (large $x$ values) will be rejected, and that at some point all jumps below a certain $x$ value are highly likely to be accepted on average. 

\begin{figure}[ht]
\centering
\includegraphics[width=0.7\textwidth]{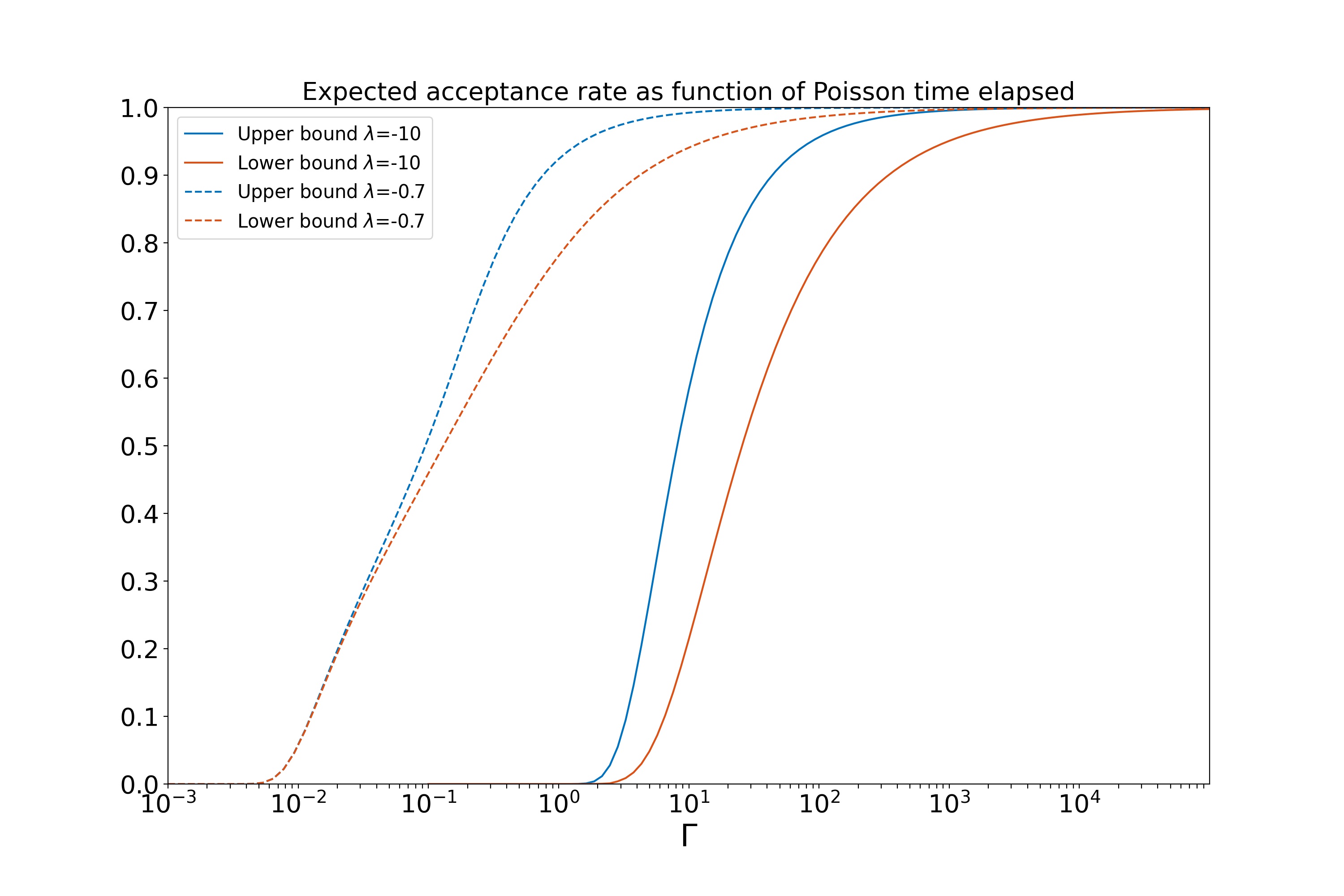}
\caption{Point acceptance intensity as a function of time evolution in the unit rate Poisson process $\{\Gamma_i\}$. $\gamma=0.2$, $\delta=0.1$.}
 \label{accept_rates}
\end{figure}

\begin{figure}[ht]
\centering
\includegraphics[width=0.7\textwidth]{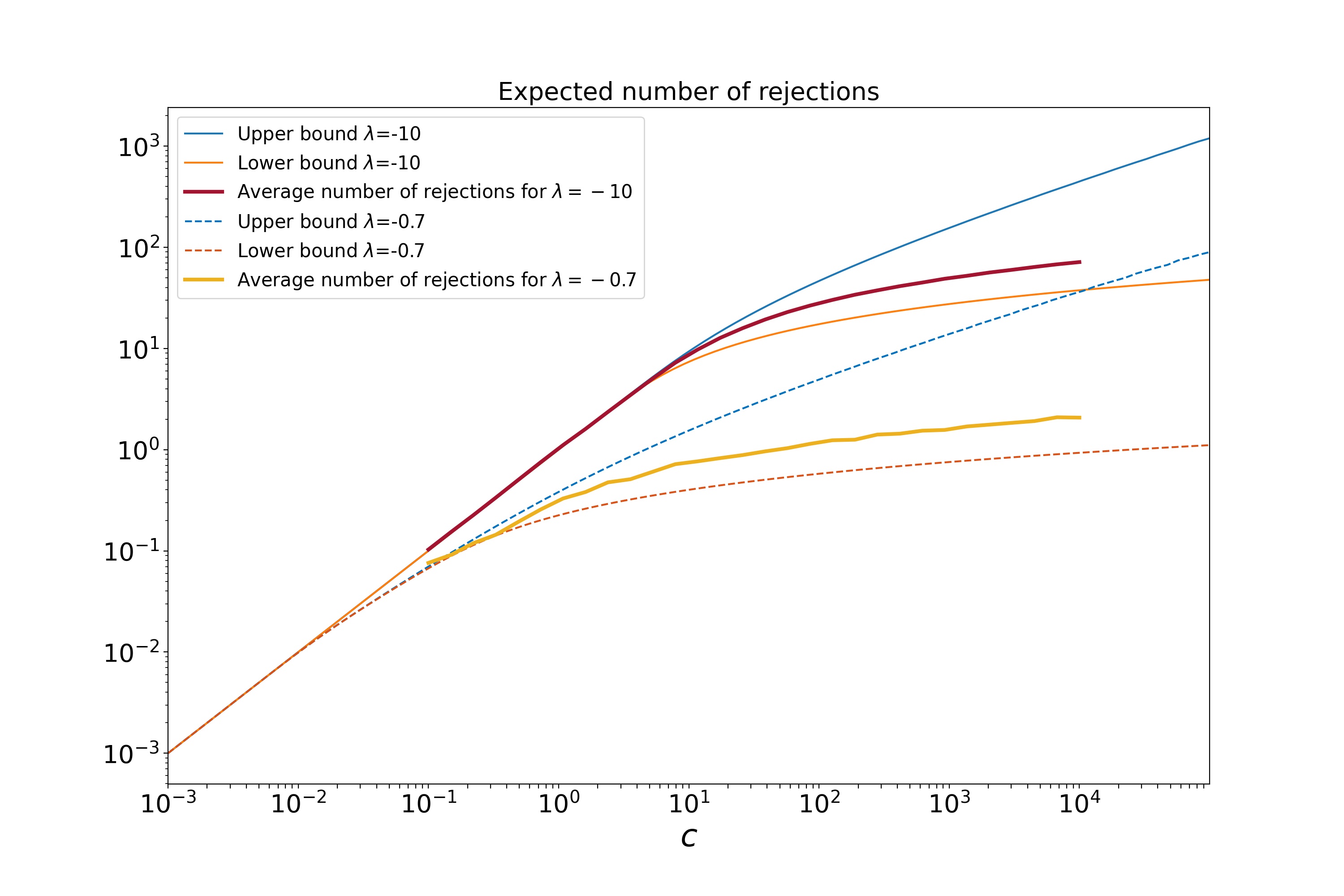}
\caption{Expected number of point process rejections as a function of the truncation level $c$ of $\{\Gamma_i\}$. $\gamma=0.2$, $\delta=0.1$, showing upper and lower bounds, plus average number of points rejected in 1000 sample paths generated by the shot noise method.} \label{rejections}
\end{figure}

In principle we can go a little further than the acceptance rate for fixed $x$ and compute overall acceptance rates for the algorithm, and quantities such as the expected number of acceptances/rejections overall. In particular it may be informative to calculate the expected number of accepted points if the underlying Poisson process $\{\Gamma_i\}$ is truncated to values of less than say $c$, i.e. the process is approximated using the random truncation $\sum_{\Gamma_i\leq c}h(\Gamma_i)$. Since $\{\Gamma_i\}$ is a unit rate Poisson process, the expected number of accepted points at a truncation level $c$ is:
\begin{equation}
N_R(c)=\int_0^c \alpha \left(h(\Gamma) \right)\rho \left(h(\Gamma) \right)d\Gamma \label{Accepted_points}
\end{equation}
Here $h(\Gamma)$ is the non-increasing function used to generate a particular point process (\ref{shot_noise_gen}), and in the case of Algorithm \ref{Q_GIG_gen} this is the tempered stable process with $h(\Gamma)=\left(\frac{\alpha\Gamma}{C}\right)^{-1/\alpha}$. $\alpha(x)$ is the acceptance probability in any accept-reject step prior to generating $z$; in this case this is the term $\exp(-\beta x)$ from the accept-reject step from the tempered stable process, where $\beta=\gamma^2/2$ (Algorithm \ref{gen_temp_stable}). Computing the integrand in (\ref{Accepted_points}) gives the point process intensity of acceptances as a function of the unit rate Poisson process's time evolution. Figure
 \ref{accept_rates}
 illustrates this by plotting the upper and lower bounds on this intensity for two values of $\lambda<-0.5$. Note that on average many fewer points are accepted at the start of the series for the larger magnitude $\lambda$, which lends some theoretical weight to the empirically observed property that the series are more slowly converging as $\lambda$ becomes increasingly negative (and the process more light-tailed), see experimental simulations section for further results on this effect. We also plot in Fig. \ref{rejections} the total number of expected rejections, computed as $c-N_R(c)$ by numerically integrating (\ref{Accepted_points}) with the upper and lower bounds substituted for the integrand. This serves to illustrate again the significant impact of $\lambda$ on total number of rejected points for a particular truncation limit $c$. Also overlaid on this figure are the actual mean number of rejections averaged over 1000 simulations of the relevant process with different truncation levels $c$, illustrating that the true expectation lies between the theoretical bounds.

\subsection{Simulation of GIG processes with $0 <|\lambda|<0.5$}

The simple bound of Theorem \ref{Lemma_1} cannot be straightforwardly applied to simulation the GIG process for  $|\lambda| <0.5$. Instead we use the more sophisticated piecewise bounds of Theorem \ref{Lemma_2} and Corollary \ref{cor_piecewise}, which give us the dominating point process (\ref{GIG_bound_lam_le_0_5}) that can be split into two independent point processes $N_1$ and $N_2$ as follows:

\begin{align*}
    N_1:\,\,\,\,\, &\frac{ e^{-x\gamma^2/2}}{\pi^2x^{1+|\lambda|}}\frac{(2\delta^2)^{|\lambda|}\gamma(|\lambda|,z_0^2x/(2\delta^2))}{H_0z_0^{2|\lambda|-1}} \frac{\Gamma(|\lambda|)\sqrt{\text{Ga}}
(z||\lambda|,x/(2\delta^2))} {\gamma(|\lambda|,z_0^2x/(2\delta^2))}{\cal I}_{z<z_0}
\end{align*}

\begin{align*}
    N_2:\,\,\,\,\, &\frac{ e^{-x\gamma^2/2}}{\pi^2x^{3/2}}\frac{(2\delta^2)^{0.5}\Gamma(0.5,z_0^2x/(2\delta^2))}{H_0} \frac{\Gamma(0.5)\sqrt{\text{Ga}}
(z|0.5,x/(2\delta^2))}{\Gamma(0.5,z_0^2x/(2\delta^2))}{\cal I}_{z\geq z_0}
\end{align*}

\noindent Each may be simulated using a thinned tempered stable process for $x$ and a truncated $\sqrt{Ga}$ density for $z$. Having simulated each pair $(x,z)$ they are accepted with probability equal to the ratio $Q_{GIG}(x,z)/Q^0_{GIG}(x,z)$. Owing to the piecewise form of the acceptance probability the two processes may be treated independently, including accept reject steps and the point process union taken at the final step to achieve the final GIG samples. The sampling procedure for point process $N_1$ is given in Algorithm \ref{gen_N_1}.

\begin{algorithm}
\caption{Generation of $N_1$}
\label{gen_N_1}
\begin{enumerate}
    \item $N_1={\emptyset}$
    \item Use the thinned series method to simulate points $x_i$, $i=1,2,3...$ from a point process with intensity function: 
    
\[
   Q_1(x)= \frac{e^{-x\gamma^2/2}}{\pi^2x^{1+|\lambda|}}\frac{(2\delta^2)^{|\lambda|}\gamma(|\lambda|,z_0^2x/(2\delta^2))}{H_0z_0^{2|\lambda|-1}}
\]

\noindent which is a modified tempered $|\lambda|$-stable process, see Section \ref{sec:marginal_x} and Algorithm \ref{Gen_Q_1}.
    \item For each $x_i$, simulate a $z_i$ from a truncated square-root gamma density:
    
\[
    \frac{\Gamma(|\lambda|)\sqrt{\text{Ga}}
(z||\lambda|,x_i/(2\delta^2))}{\gamma(|\lambda|,z_0^2x_i/(2\delta^2))}{\cal I}_{0<z<z_0}
\]

\item  With probability    
    \[
    \frac{H_0}{|H_{|\lambda|}(z_i)|^2\left(\frac{z_i^{2|\lambda|}}{z_0^{2|\lambda|-1}}
\right)}
\]

\noindent accept $x_i$, i.e. set $N_1=N_1\cup x_i$, otherwise discard $x_i$.
\end{enumerate}
\end{algorithm}

Similarly, for $N_2$, the procedure is given in Algorithm \ref{gen_N_2}.

\begin{algorithm}
\caption{Generation of $N_2$}
\label{gen_N_2}
\begin{enumerate}
    \item $N_2={\emptyset}$
    \item Use the thinned series method to simulate points $x_i$, $i=1,2,3...$ from a point process with intensity function:  
\[
    Q_2(x)=\frac{e^{-x\gamma^2/2}}{\pi^2x^{3/2}}\frac{(2\delta^2)^{0.5}\Gamma(0.5,z_0^2x/(2\delta^2))}{H_0}
\] using Algorithm \ref{Gen_N_2}.
    \item For each $x_i$, simulate a $z_i$ from a truncated square-root gamma density:
    
\[
    \frac{\Gamma(0.5)\sqrt{\text{Ga}} (z|0.5,x_i/(2\delta^2))}{\Gamma(0.5,z_0^2x_i/(2\delta^2))}{\cal I}_{z\geq z_0}
\]
\item  With probability:

\[
    \frac{H_0}{z_i|H_{|\lambda|}(z_i)|^2}
\]

\noindent accept  $x_i$, i.e. set $N_2=N_2\cup x_i$, otherwise reject $x_i$
\end{enumerate}

\end{algorithm}

Finally, the set of points $N=N_1\cup N_2$ is a realisation of jump sizes from the point process having intensity $Q_{GIG}(x)$. The procedure is completed as before by generating independent jump times $v\sim {\cal U}[0,T]$ for all points $x\in N$. 

\subsubsection{Acceptance Rates}


 Theorem \ref{Lemma_2} is used once again to find lower bounds on the expected acceptance rates. In this case the bound to apply is $z|H_{\lambda}(z)|^2 \leq A(z)$.


For $N_1$ the average acceptance rate $\rho_1(x)$ is
\begin{align*}
    \rho_1(x) &= E \left[ \frac{H_0}{|H_{|\lambda|}(z)|^2 \left( \frac{z^{2 |\lambda| }}{z_0^{2|\lambda|-1}}\right)} \right] \\ &= \int_0^{z_0}\frac{\Gamma(|\lambda|)\sqrt{\text{Ga}}
(z||\lambda|,x/(2\delta^2))}{\gamma(|\lambda|,z_0^2x/(2\delta^2))} \\
& \quad \quad \quad \frac{H_0}{z|H_{|\lambda|}(z)|^2} \left(\frac{z_0}{z}\right)^{2|\lambda|-1}
dz
\end{align*}
which may be lower bounded by substituting the bound $z|H_{\lambda}(z)|^2 \leq A(z)$ from Theorem \ref{Lemma_2} and by direct integration to give: 
\[
    \rho_1(x) \geq \frac{H_0 z_0^{2|\lambda|-1} \pi^2}{2^{2|\lambda|} \Gamma^2(|\lambda|)}
\]

Similarly, for $N_2$ the average acceptance rate $\rho_2(x)$ is

\begin{align*}
 \rho_2(x) &= E \left[ \frac{H_0}{z|H_{|\lambda|}(z)|^2} \right] \\ &= \int_{z_0}^{\infty} \frac{\Gamma(0.5)\sqrt{\text{Ga}} (z|0.5,x/(2\delta^2))}{\Gamma(0.5,z_0^2x/(2\delta^2))} \frac{H_0}{z|H_{|\lambda|}(z)|^2} dz
\end{align*}
and again lower bounds are obtained by substituting $z|H_{\lambda}(z)|^2 \leq A(z)$ from Theorem \ref{Lemma_2} and direct integration to give:
\[
\rho_2(x) \geq \frac{\pi H_0}{2}
\]

We note that both of these bounds are in fact independent of the value of $x$. If we use $B(z)$ to obtain an upper bound on the acceptance rates $\rho_1$ and $\rho_2$, both bounds are found to be $1$, i.e. no useful information is gleaned from the upper bound.

\subsection{Sampling from the marginal point process envelope}
\label{sec:marginal_x}

In steps 2 above we require simulation of the marginal point process for $x$ in the dominating bivariate point process. For $N_1$ this has intensity:

\begin{equation}
\begin{aligned}
    Q_1(x) &= \frac{e^{-x\gamma^2/2}}{\pi^2x^{1+|\lambda|}}\frac{(2\delta^2)^{|\lambda|}\gamma(|\lambda|,z_0^2x/(2\delta^2))}{H_0z_0^{2|\lambda|-1}} \\ &= \frac{e^{-x\gamma^2/2}}{\pi^2x}\frac{\gamma(|\lambda|,z_0^2x/(2\delta^2))}{(z_0^2 x/(2\delta^2))^{|\lambda|} H_0 z_0^{-1}}
\end{aligned}
\end{equation}

\noindent and we propose two possible methods for simulating this point process.

The first and most basic method uses the fact that the lower incomplete gamma function is upper bounded by the complete gamma function, i.e. $\gamma(|\lambda|,z_0^2x/(2\delta^2))\leq \Gamma(|\lambda|)$, which allows generation from $Q_1$ by thinning of a tempered stable process. We thus have the following upper bounding envelope as a tempered stable density:

\[
    Q_1(x)\leq \bar{Q}_1^1(x) = \frac{e^{-x \gamma^2/2} \Gamma(|\lambda|)}{\pi^2 x^{1+|\lambda|}} \frac{(2\delta^2)^{|\lambda|}}{H_0 z_0^{2|\lambda|-1}}
\]

\noindent This process may be routinely sampled by thinning of a positive tempered $|\lambda|$-stable process as described in \ref{TS_sample}.

Having generated points from the tempered stable envelope function the  process is thinned with probability:

\[
    \frac{Q_1(x)}{\bar{Q}_1^1(x)}=\frac{\gamma(|\lambda|,z_0^2x/(2\delta^2))}{\Gamma(|\lambda|)}
\]

\noindent Experimentally we found that the acceptance rates were quite low for this bound, meaning that fairly long tempered stable series had to be generated (but see note below about the case $\gamma=0$), so a more sophisticated bound was sought, as follows.

The second and more effective method employs the following bound on the lower incomplete gamma function (\cite{Neuman2013}, Theorem 4.1):
\[
    \frac{a\gamma(a,x)}{x^a} \leq \frac{(1+ae^{-x})}{(a+1)}
\]
and so
\begin{equation}
    \frac{\gamma(|\lambda|,z_0^2 x/(2\delta^2))}{(z_0^2 x/(2\delta^2))^{|\lambda|}} \leq \frac{(1+|\lambda|e^{-z_0^2 x/(2\delta^2)})}{|\lambda|(|\lambda|+1)}
\end{equation}

\noindent so that the point process intensity function is upper bounded by

\begin{align}
Q_1(x)&\leq \frac{e^{-x\gamma^2/2}}{\pi^2x}\frac{z_0(1+|\lambda|e^{-z_0^2 x/(2\delta^2)})}{|\lambda|(1+|\lambda|)H_0}={\bar{Q}_1^{2a}(x)}\label{Q_1_intermediate}\\
&\,\,\,\,\,\leq \frac{e^{-x\gamma^2/2}}{\pi^2x}\frac{z_0}{|\lambda|H_0}={\bar{Q}_1^2(x)} \label{Q_1_2_bar}
\end{align}

\noindent where (\ref{Q_1_2_bar}) follows from $0<e^{-z_0^2 x/(2\delta^2)} \leq 1$ for $x \in [0, \infty)$. The bounding process $\bar{Q}^2_1(x)$ may be simulated as a single gamma process, having parameters
$a = \frac{z_0}{\pi^2 |\lambda|H_0}$ and $\beta=\gamma^2/2$. It is then thinned with the following probability:
\begin{align}
    \frac{Q_1(x)}{\bar{Q}_1^2(x)} &= \frac{|\lambda| \gamma(|\lambda|, z_0^2 x/(2\delta^2))}{\left(\frac{z_0^2 x}{2 \delta^2} \right)^{|\lambda|}}\label{N_1_thin} 
    \end{align}
The whole procedure is given in Algorithm \ref{Gen_Q_1}.

\begin{algorithm}
\caption{Generating from $Q_1$}
\label{Gen_Q_1}
\begin{enumerate}
    \item Generate a gamma process $N_{MGa}$ using Algorithm \ref{gen_gamma} having parameters
$a = \frac{z_0}{\pi^2 |\lambda|H_0}$ and $\beta=\gamma^2/2$
\item For each point $x\in N_{MGa}$, accept with probability (\ref{N_1_thin}), otherwise reject and delete $x$ from $N_{MGa}$.
\end{enumerate}
\end{algorithm}

Note that a slightly more refined bounding procedure may be implemented by observing that (\ref{Q_1_intermediate}) is the intensity function of the union of two gamma processes, which may be independently simulated and their union thinned with probability $\frac{Q_1(x)}{\bar{Q}_1^{2a}(x)}$, at the expense of generating one additional gamma process.

Note also that the second method using Algorithm \ref{Gen_Q_1} does not work for $\gamma=0$ since the bounding point process cannot be simulated in this case; instead the first method with bound $\bar{Q}_1^1(x)$ must be used for the $\gamma=0$ case.
This second method, using bounding function $\bar{Q}_1^2(x)$, is regarded as the superior approach for all parameter settings except for $\gamma=0$ since it has increased acceptance probabilities and also relies on an underpinning Gamma process rather than a $|\lambda|$-stable process as its basis, and the series representation of the Gamma process is known to converge very rapidly to zero compared to the $|\lambda|$-stable process \cite{Rosinski_2001,B_N_Shephard_2001}.

Similarly, for $N_2$ we  follow a thinning approach. The marginal dominating L\'{e}vy density for $N_2$ is:

\begin{equation}
    Q_2(x) = \frac{e^{-x\gamma^2/2}}{\pi^2x^{3/2}}\frac{(2\delta^2)^{0.5}\Gamma(0.5,z_0^2x/(2\delta^2))}{H_0}
\end{equation}

\noindent Using the complete gamma function as part of an upper bound for this density, we can sample $N_2$ as a tempered stable intensity $\frac{e^{-x\gamma^2/2} (2\delta^2)^{0.5} \Gamma(0.5)}{\pi^2H_0 x^{3/2}}$, and apply thinning with probability $\Gamma(0.5, z_0^2 x/(2\delta^2))/\Gamma(0.5)$. This procedure is found to work well and the procedure is summarised in 
Algorithm \ref{Gen_N_2}. 
\begin{algorithm}
\caption{Generating from $Q_2$}
\label{Gen_N_2}
\begin{enumerate}
    \item Generate a tempered stable process $N_{MTS}$ using Algorithm \ref{gen_temp_stable} with parameters $C=\frac{(2\delta^2)^{0.5} \Gamma(0.5)}{\pi^2H_0}$, $\alpha=0.5$ and $\beta=\gamma^2/2$. 
\item For each point $x\in N_{MTS}$, accept with probability $\Gamma(0.5, z_0^2 x/(2\delta^2))/\Gamma(0.5)$, otherwise reject and delete $x$ from $N_{MTS}$.
\end{enumerate}
\end{algorithm}

Rejection rates could potentially be further  improved by more tightly upper bounding the term $\Gamma(0.5, \frac{z_0^2 x}{2\delta^2})$. The following simple bound is suitable, and valid for $0<s<1$:
\[
    \Gamma(s,x) \leq x^{s-1} e^{-x} 
\]
This was not explored in the simulations as Algorithm \ref{Gen_N_2} was found to perform well.



\section{Simulations}

\begin{figure}[tbp]
  \includegraphics[width=0.9\textwidth]{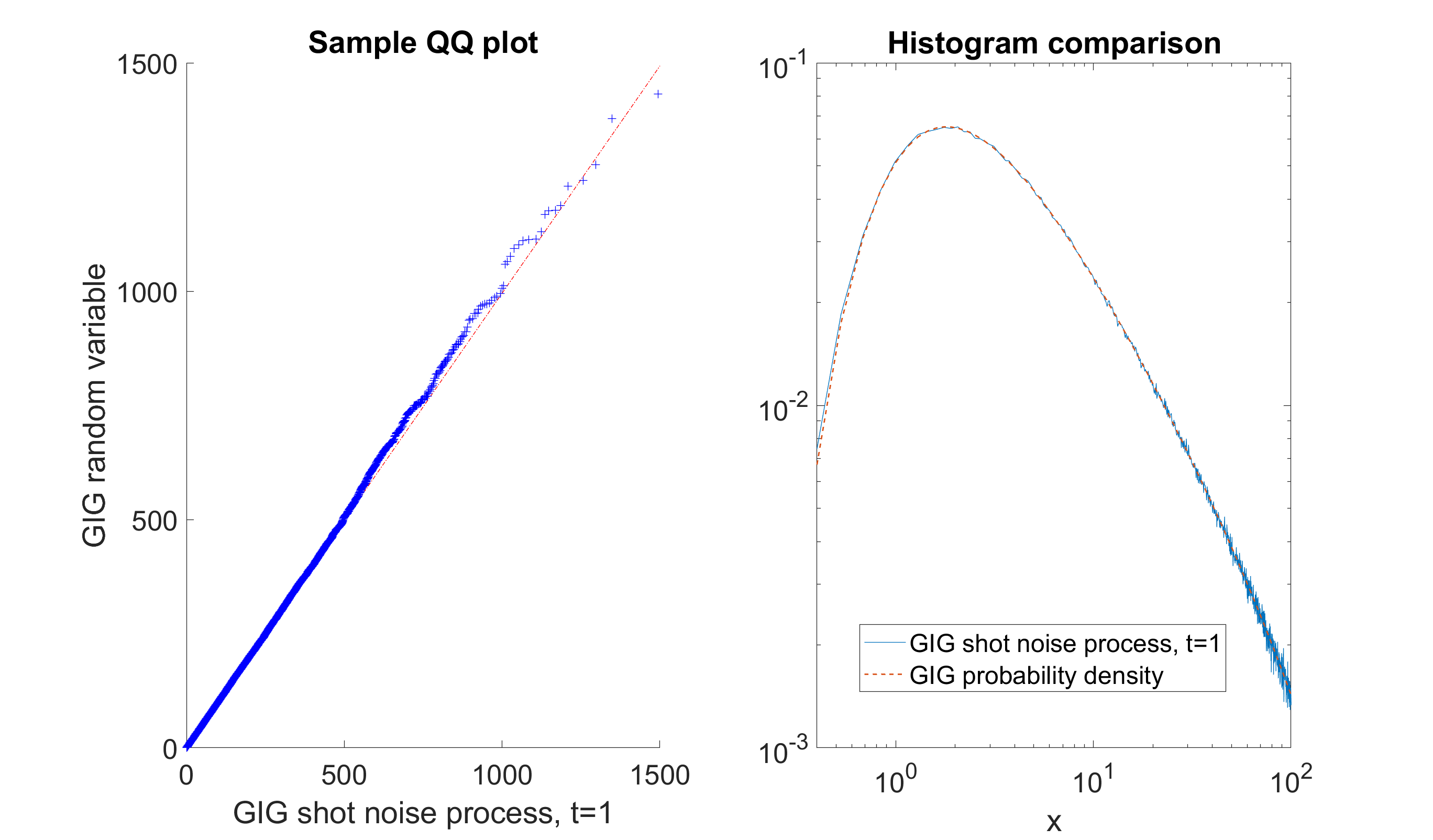}
\caption{Simulation comparison between the shot noise generated GIG process and GIG random variates, $\lambda=-0.1$, $\gamma=0.1$, $\delta=2$, $10^6$ random samples.Left hand panel: QQ plot comparing our shot noise method (x-axis) with random samples of the GIG density generated using the `ground truth' method of \cite{Devroye_2014,Statovic_2017}. Right hand panel: Normalised histogram density estimate  for our method compared with the true GIG density function.}
\label{hist_qq_m0_1}       
\end{figure}

\begin{figure}[tbp]
  \includegraphics[width=0.9\textwidth]{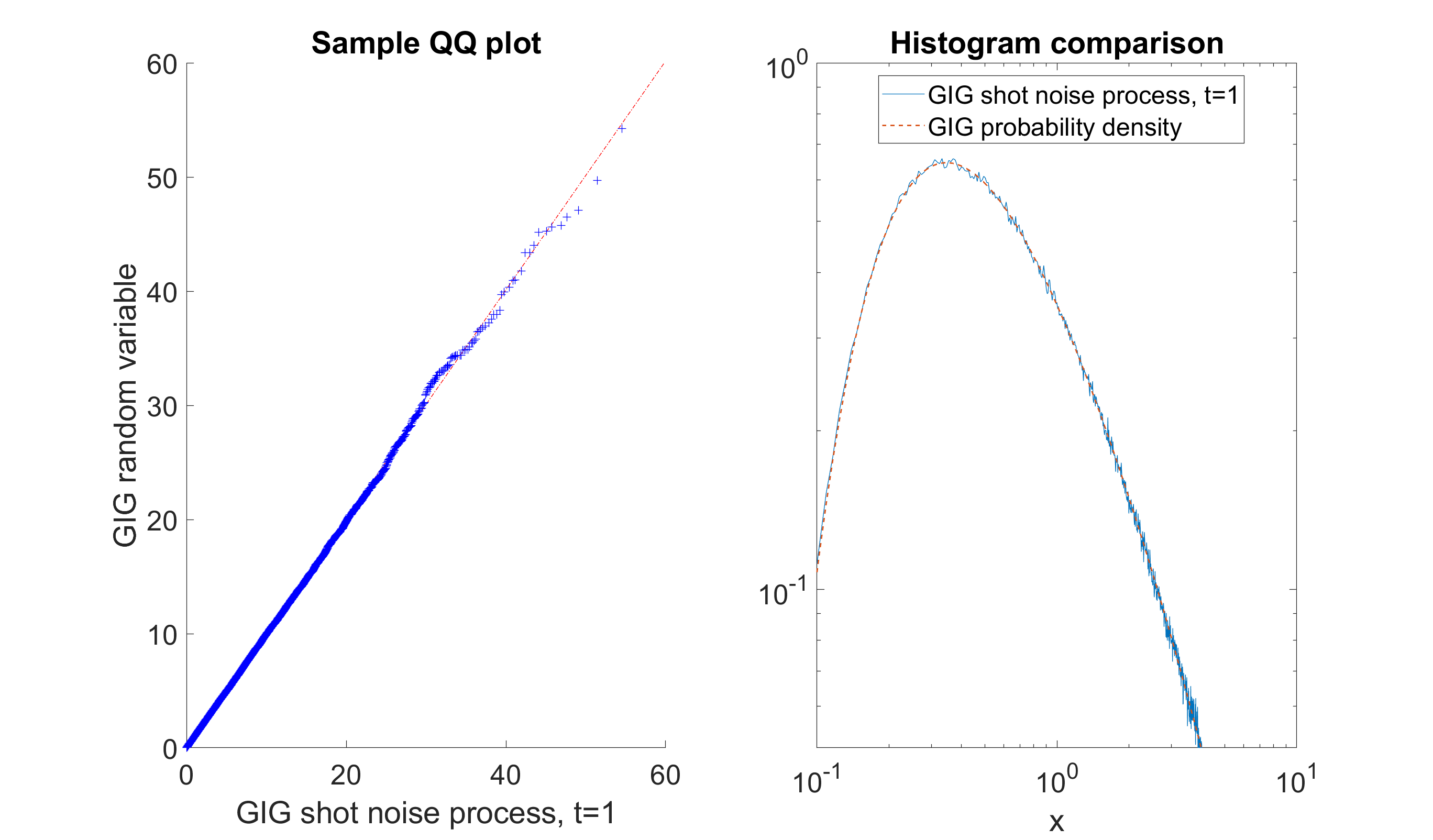}
\caption{Simulation comparison between the shot noise generated GIG process and GIG random variates, $\lambda=-0.4$, $\gamma=0.5$, $\delta=1$, $10^6$ random samples. Left hand panel: QQ plot comparing our shot noise method (x-axis) with random samples of the GIG density generated using the `ground truth' method of \cite{Devroye_2014,Statovic_2017}. Right hand panel: Normalised histogram density estimate  for our method compared with the true GIG density function.}
\label{hist_qq_m0_4}       
\end{figure}

\begin{figure}[tbp]
\centering
\includegraphics[width=0.9\textwidth]{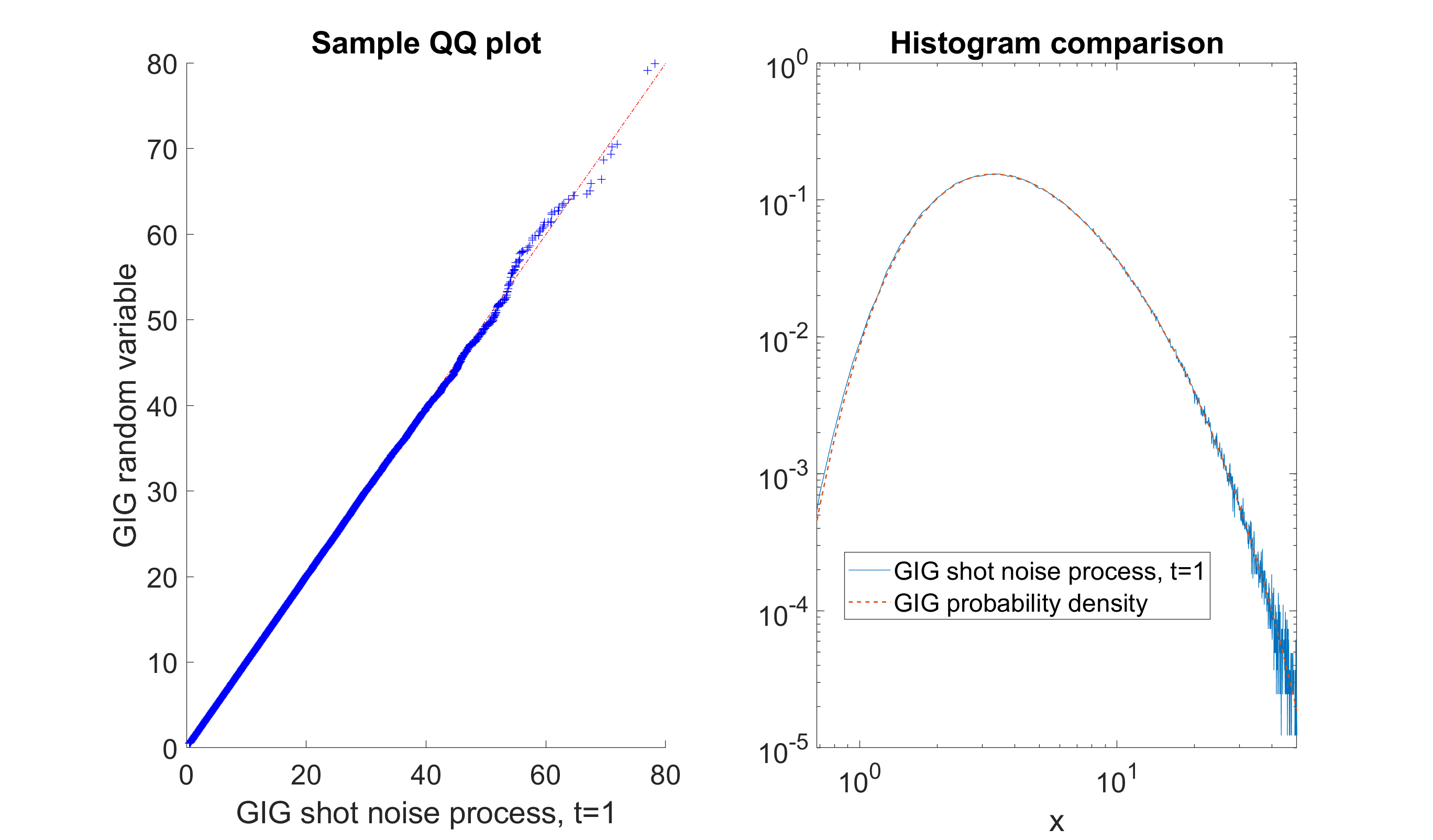}
\caption{Simulation comparison between the shot noise generated GIG process and GIG random variates, $\lambda=-1$, $\gamma=0.5$, $\delta=4$, $10^6$ random samples. Left hand panel: QQ plot comparing our shot noise method (x-axis) with random samples of the GIG density generated using the `ground truth' method of \cite{Devroye_2014,Statovic_2017}. Right hand panel: Normalised histogram density estimate  for our method compared with the true GIG density function.}
 \label{hist_qq_m1}
 \includegraphics[width=0.9\textwidth]{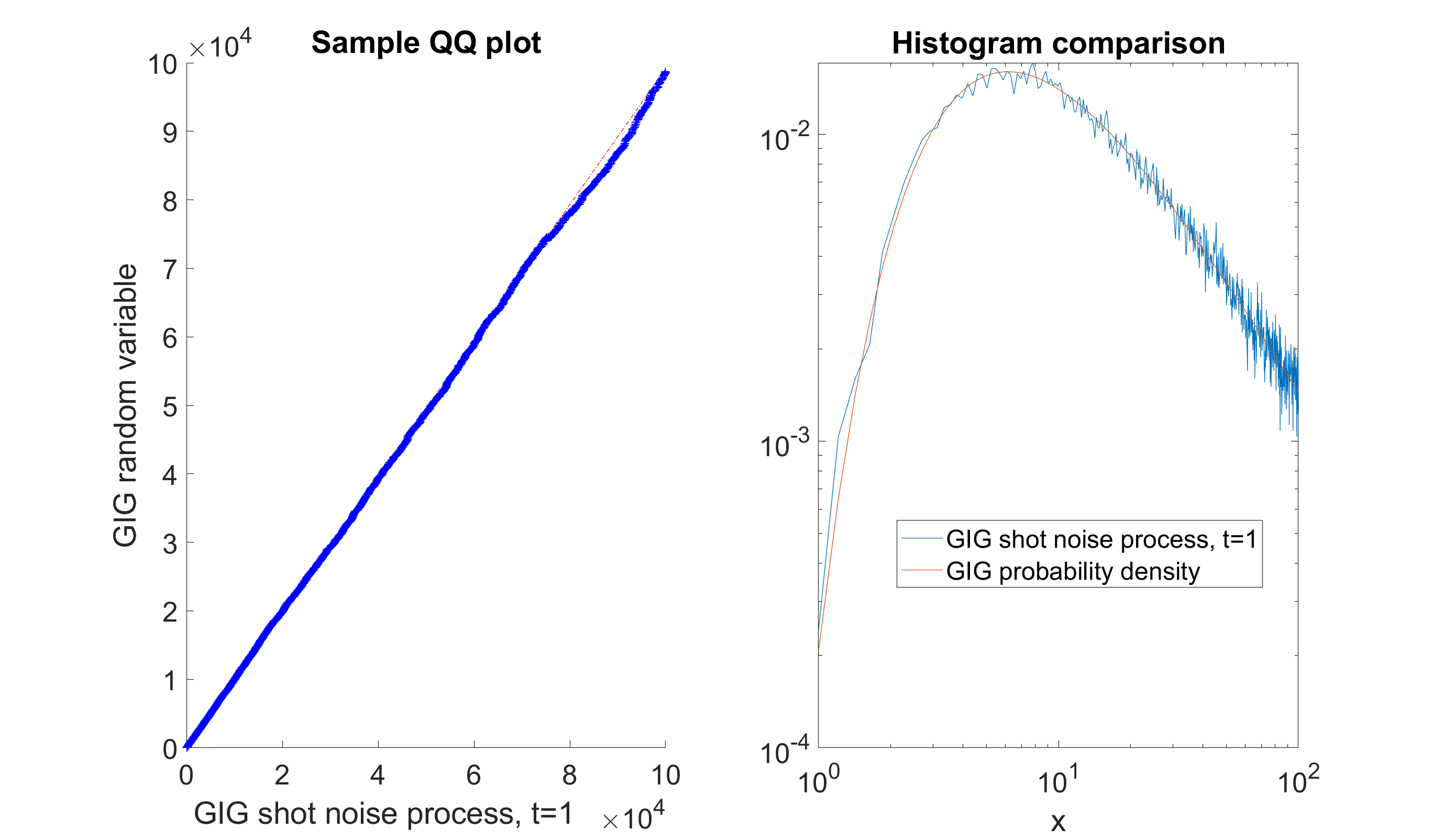}
\caption{Simulation comparison between the shot noise generated GIG process and GIG random variates, $\lambda=-0.3$, $\gamma=0$, $\delta=4$, $10^6$ random samples. Left hand panel: QQ plot comparing our shot noise method (x-axis) with random samples of the GIG density generated using the `ground truth' method of \cite{Devroye_2014,Statovic_2017}. Right hand panel: Normalised histogram density estimate  for our method compared with the true GIG density function.}
 \label{hist_qq_m0_3_gamma_0}
\end{figure}

\begin{figure}[tbp]
\centering
\includegraphics[width=0.9\textwidth]{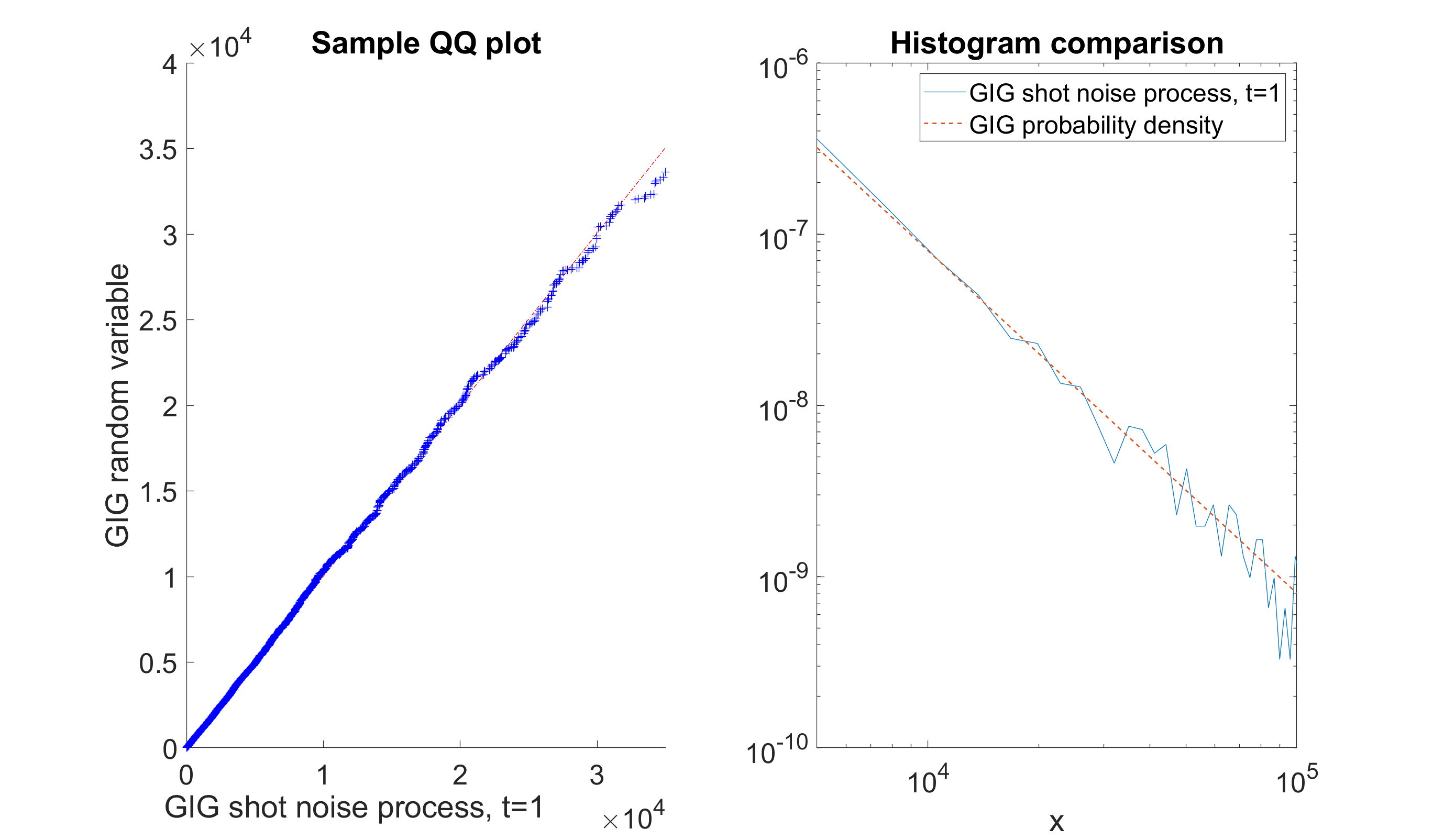}
\caption{Simulation comparison between the shot noise generated GIG process and GIG random variates, $\lambda=-1$, $\gamma=0$, $\delta=4$, $10^6$ random samples. Left hand panel: QQ plot comparing our shot noise method (x-axis) with random samples of the GIG density generated using the `ground truth' method of \cite{Devroye_2014,Statovic_2017}. Right hand panel: Normalised histogram density estimate  for our method compared with the true GIG density function.}
 \label{hist_qq_m1_gamma_0}
\end{figure}

\begin{figure}[tbp]
\centering
\includegraphics[width=0.9\textwidth]{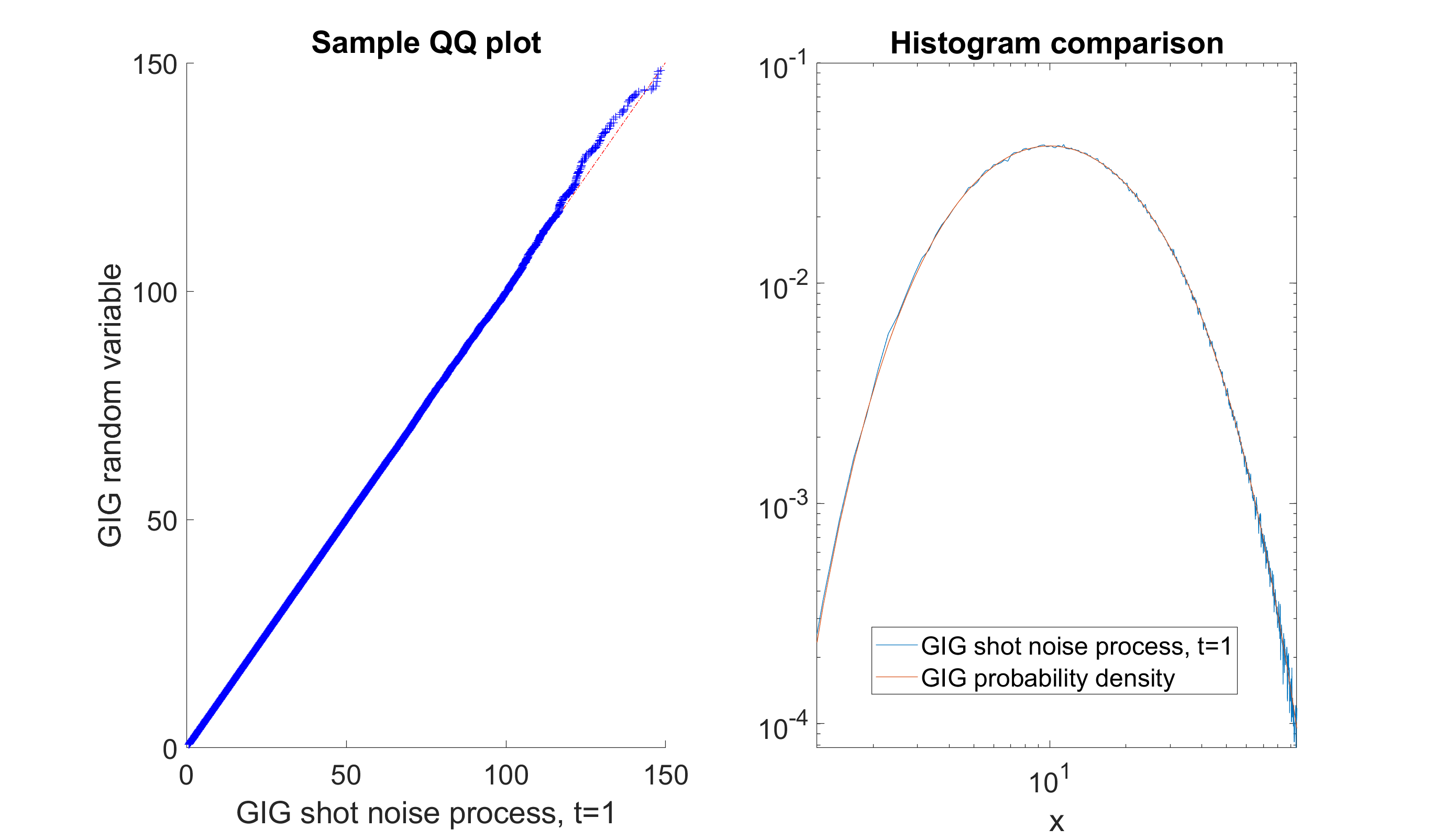}
\caption{Simulation comparison between the shot noise generated GIG process and GIG random variates, $\lambda=1$, $\gamma=0.4$, $\delta=4$, $10^6$ random samples. Left hand panel: QQ plot comparing our shot noise method (x-axis) with random samples of the GIG density generated using the `ground truth' method of \cite{Devroye_2014,Statovic_2017}. Right hand panel: Normalised histogram density estimate  for our method compared with the true GIG density function.}
 \label{hist_qq_1}
 \includegraphics[width=0.9\textwidth]{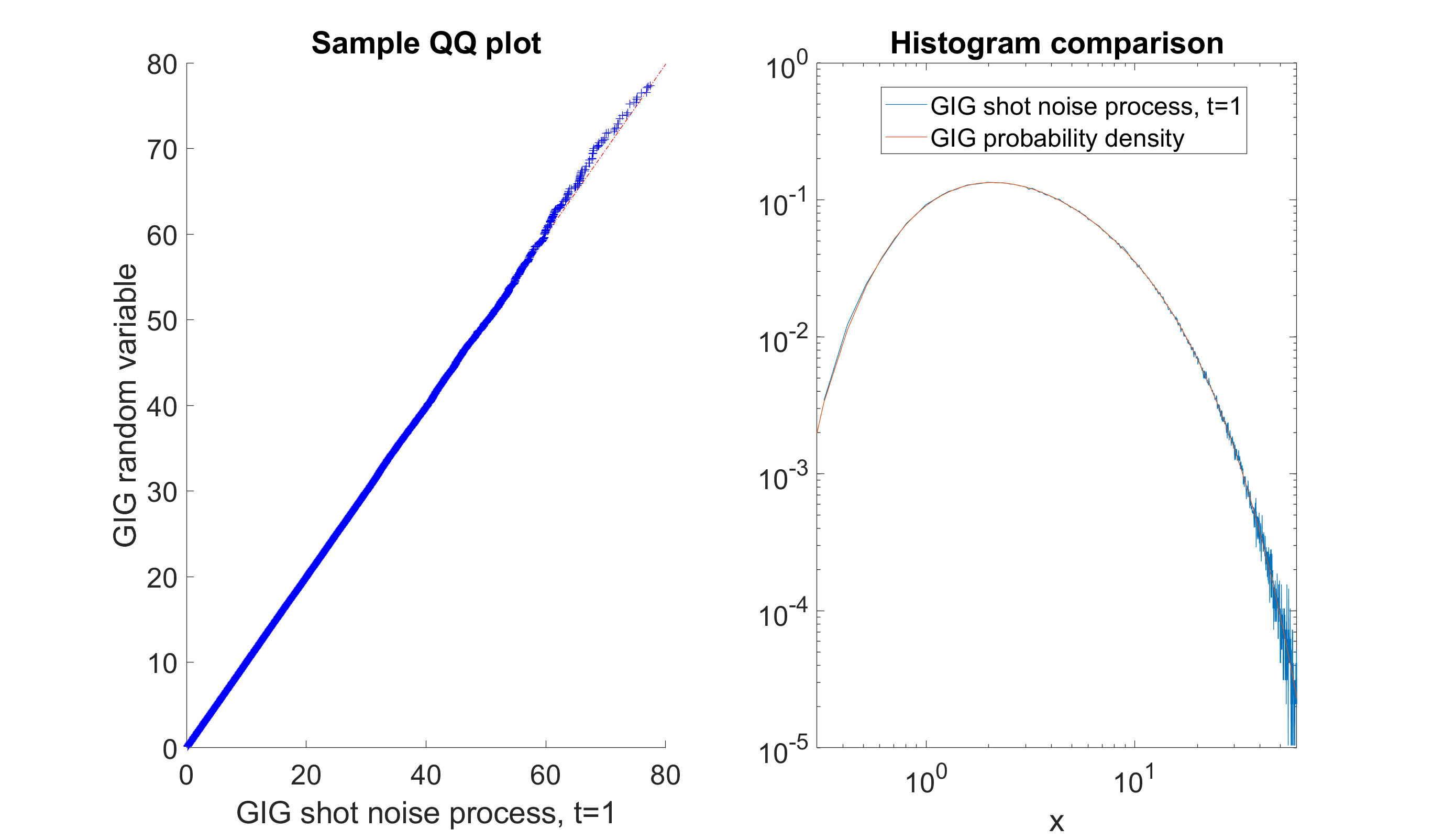}
\caption{Simulation comparison between the shot noise generated GIG process and GIG random variates, $\lambda=0.3$, $\gamma=0.5$, $\delta=2$, $10^6$ random samples. Left hand panel: QQ plot comparing our shot noise method (x-axis) with random samples of the GIG density generated using the `ground truth' method of \cite{Devroye_2014,Statovic_2017}. Right hand panel: Normalised histogram density estimate  for our method compared with the true GIG density function.}
 \label{hist_qq_0_3}
\end{figure}

\begin{figure}[tbp]
\centering
\includegraphics[width=0.9\textwidth]{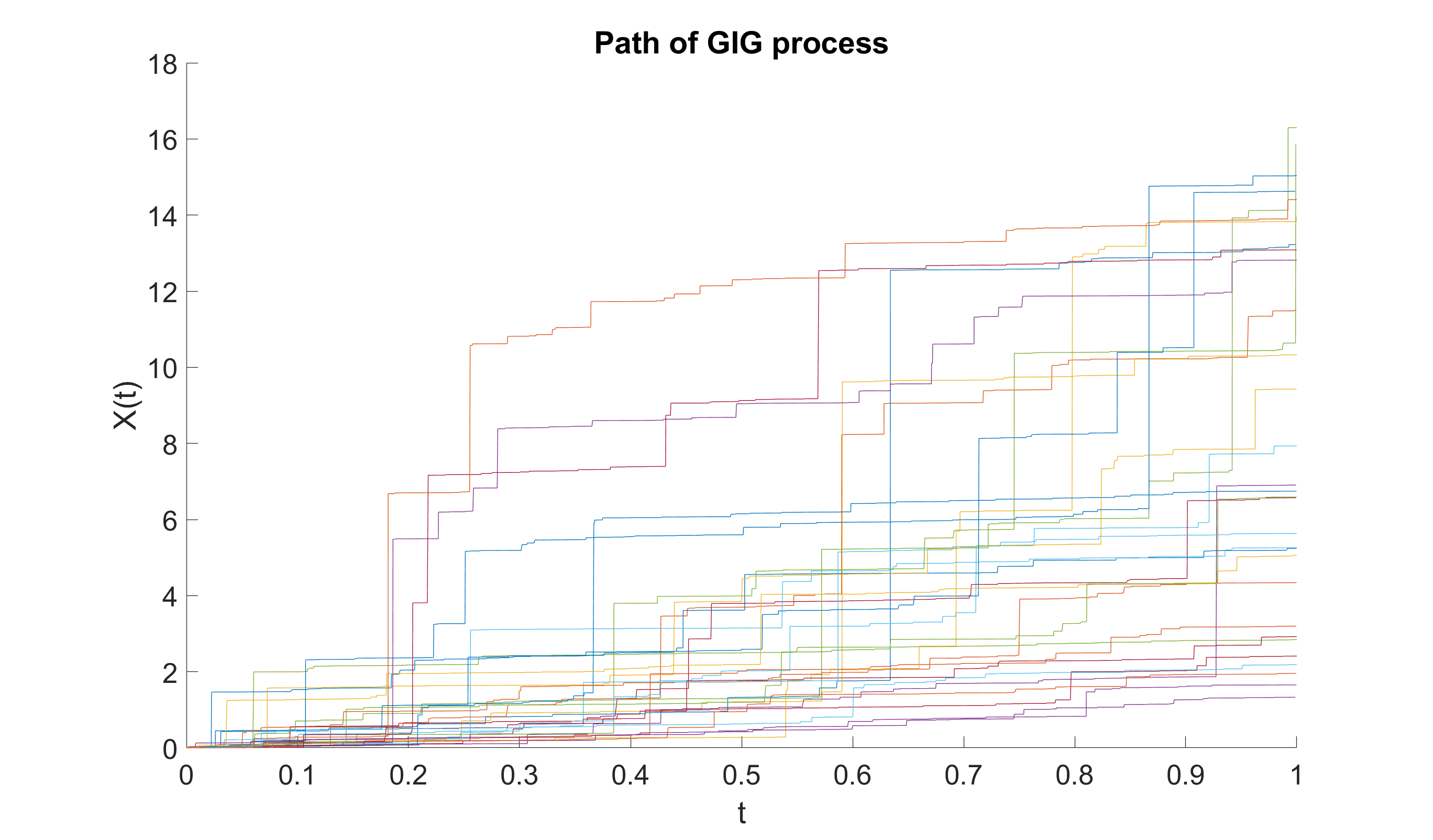}
\caption{Pathwise simulations of the GIG process, $\lambda=-0.1$, $\gamma=0.5$, $\delta=3$.}
 \label{GIG_path_m0_1}
 \includegraphics[width=0.9\textwidth]{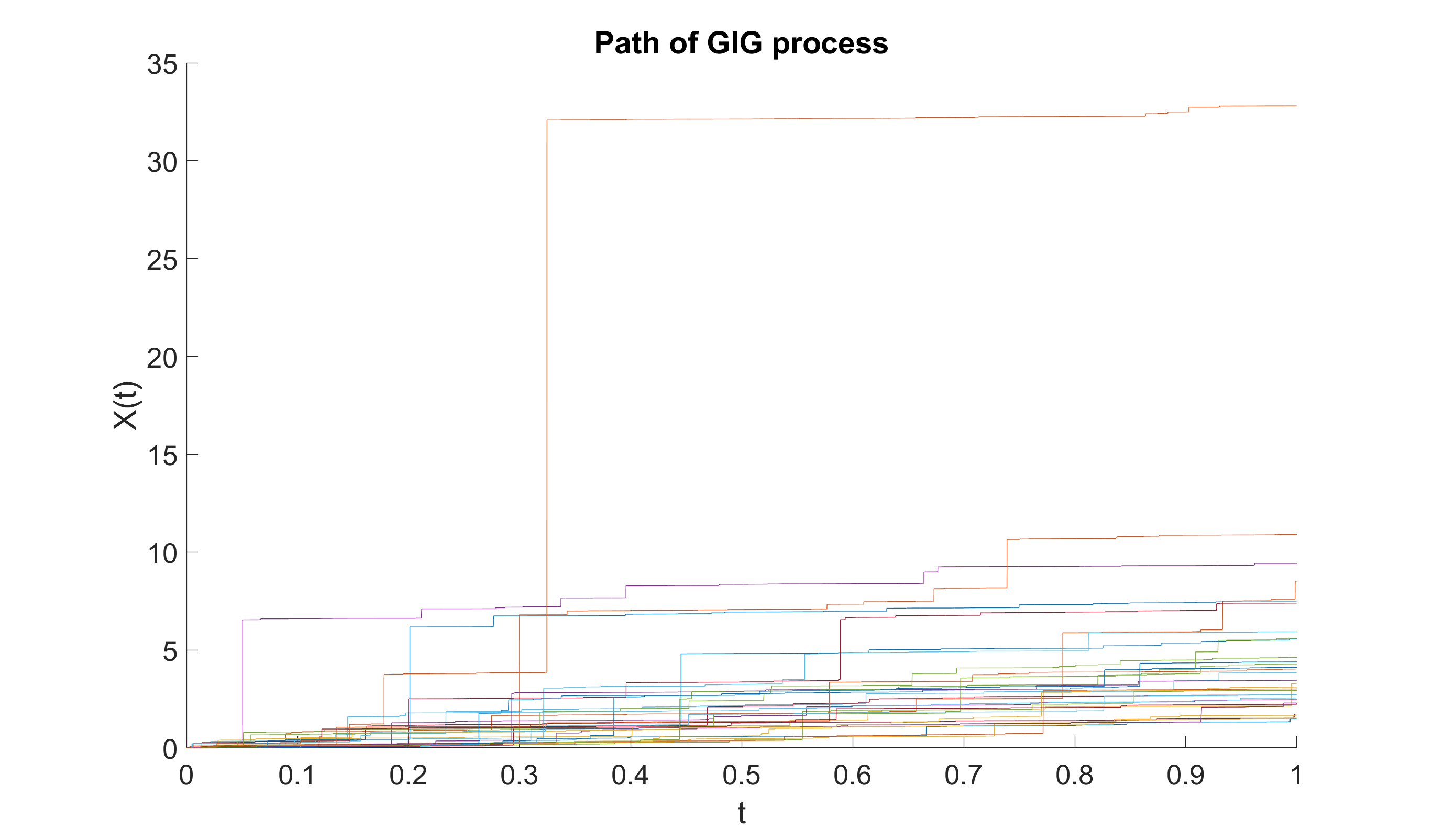}
\caption{Pathwise simulations of the GIG process, $\lambda=-1$, $\gamma=0.5$, $\delta=3$ (log-scale). }
 \label{GIG_path_m_1}
\end{figure}

\begin{figure}[tbp]
\centering
\includegraphics[width=0.9\textwidth]{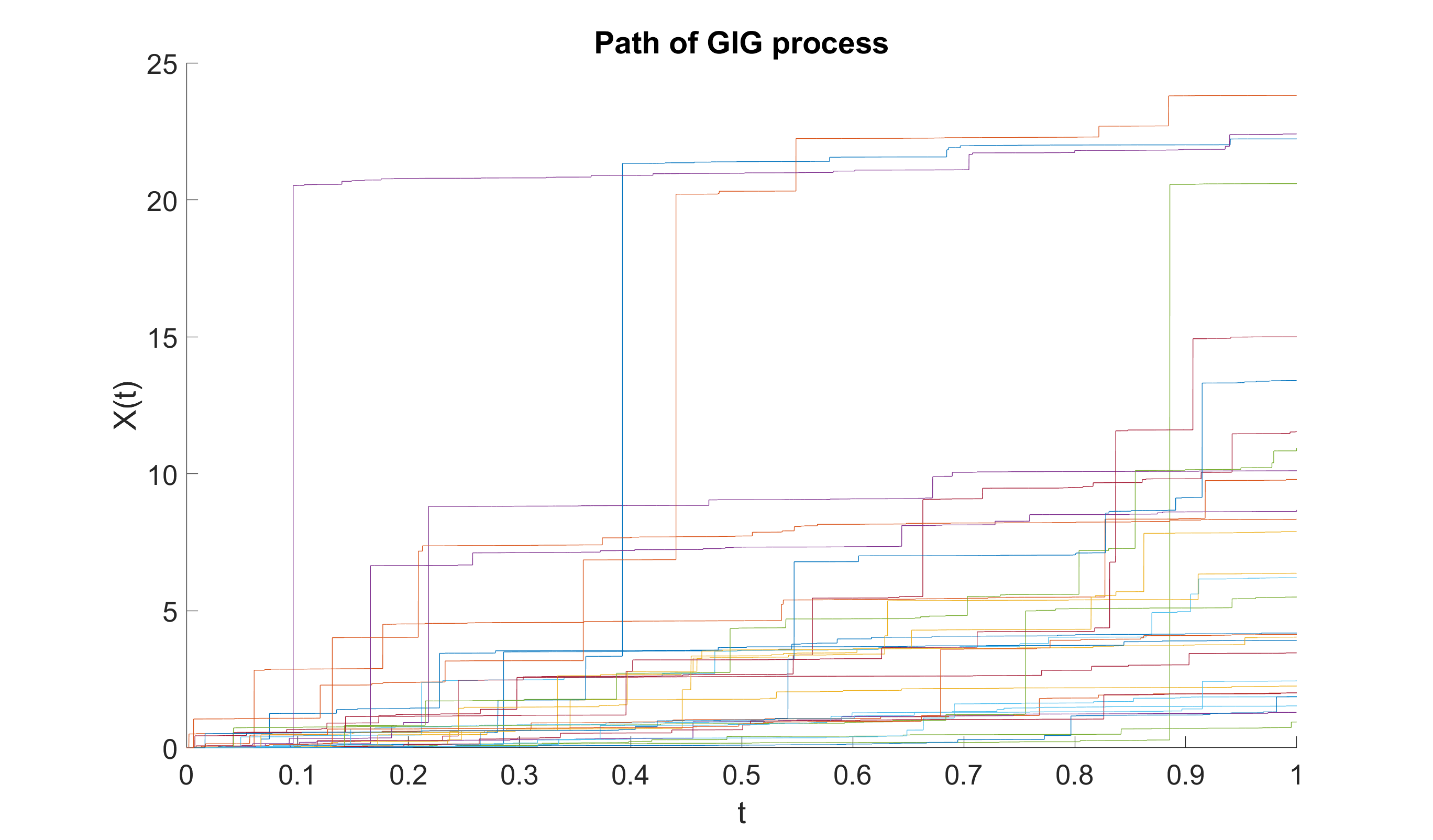}
\caption{Pathwise simulations of the GIG process, $\lambda=1$, $\gamma=0.4$, $\delta=4$.}
 \label{GIG_path_1}
 \includegraphics[width=0.9\textwidth]{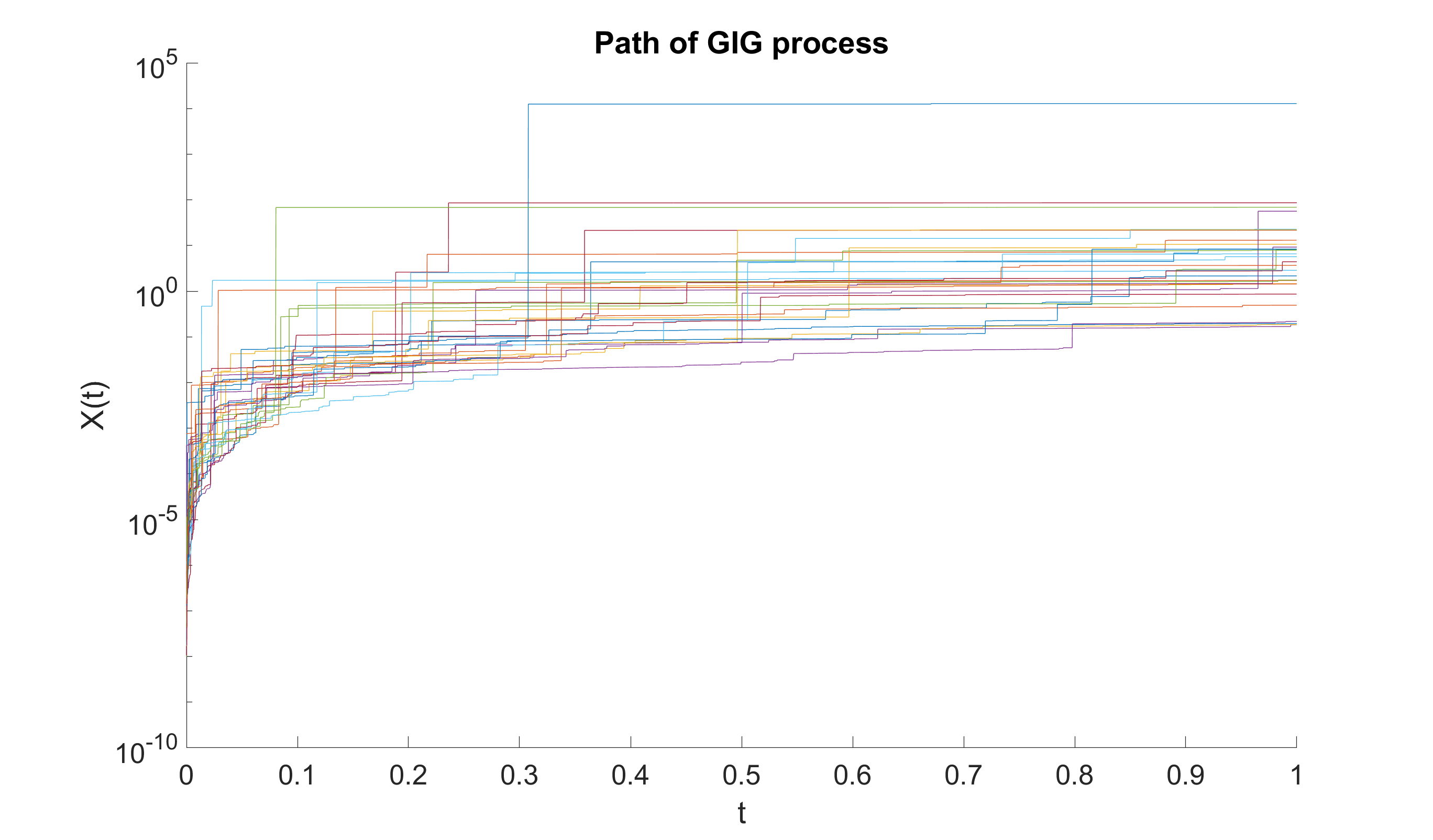}
\caption{Pathwise simulations of the GIG process, $\lambda=-0.4$, $\gamma=0$, $\delta=1$ (log-scale).}
 \label{GIG_path_m0_4_gamma_0}
\end{figure}

In this section example simulations from the new method are generated up to $t=T$, where $T=1$, and compared with a random variable generator for the GIG distribution \cite{Devroye_2014,Statovic_2017} (a `ground truth' simulation for comparing the distribution at $t=T$, but unlike our method, not able to generate the entire path of the process). In our new shot noise case, $M=1000$ terms are generated for each point process series representation and $10^6$ random realisations of the GIG random variable are generated to produce Figs. \ref{hist_qq_m0_1}-\ref{hist_qq_0_3}, in which examples of the principal parameter ranges and edge case $\gamma=0$ are presented. Note that $M$ represents the total number of terms generated for each of the underlying tempered stable and gamma processes, rather than the final number of accepted terms. $10^4$ random samples were generated in each case for our new shot noise method and $10^6$  samples for the `ground truth' method of \cite{Devroye_2014,Statovic_2017} at $t=1$. In Figs. \ref{GIG_path_m0_1}-\ref{GIG_path_m0_4_gamma_0} example pathwise simulations are plotted for several parameter settings, drawing 30 independent realisations of the GIG process in each case, setting $T=1$. For cases where $\lambda>0$ the additional term in the L\'{e}vy density, $\lambda\frac{e^{-x\gamma^2/2}}{x}$, see (\ref{GIG_levy_density}), is generated as an independent Gamma process and the union of the these points with those from $Q_{GIG}$ is taken to generate the final simulated process. Once the union of all the points has been taken, to form a set of jump sizes $\{x_i\}$, independent and uniformly distributed jump times $\{v_i\in[0,T]\}$ are generated and the process path may be realised as
\[\sum_i x_i{\cal I}_{v_i\leq t},\,\,\,t\in[0,T]\]

As previously discussed, the resulting process approximates a GIG L\'{e}vy process. By comparing the distribution of the L\'{e}vy process at $t=1$ and the distribution of GIG random variables generated using \cite{Devroye_2014,Statovic_2017}, the quality of this approximation may be evaluated, at least at the end-point of the interval $t=T=1$. A possible statistical test in this case is the two-sample Kolmogorov-Smirnov (KS) test since in most cases the CDF of a GIG random variable is not available in closed form. The KS test checks whether the two samples come from the same distribution by reporting the KS statistic, computed as the supremum of the absolute difference between the empirical CDFs of the two samples \cite{Hodges1958TheSP}.

\begin{figure}[tbp]
  \includegraphics[width=0.9\textwidth]{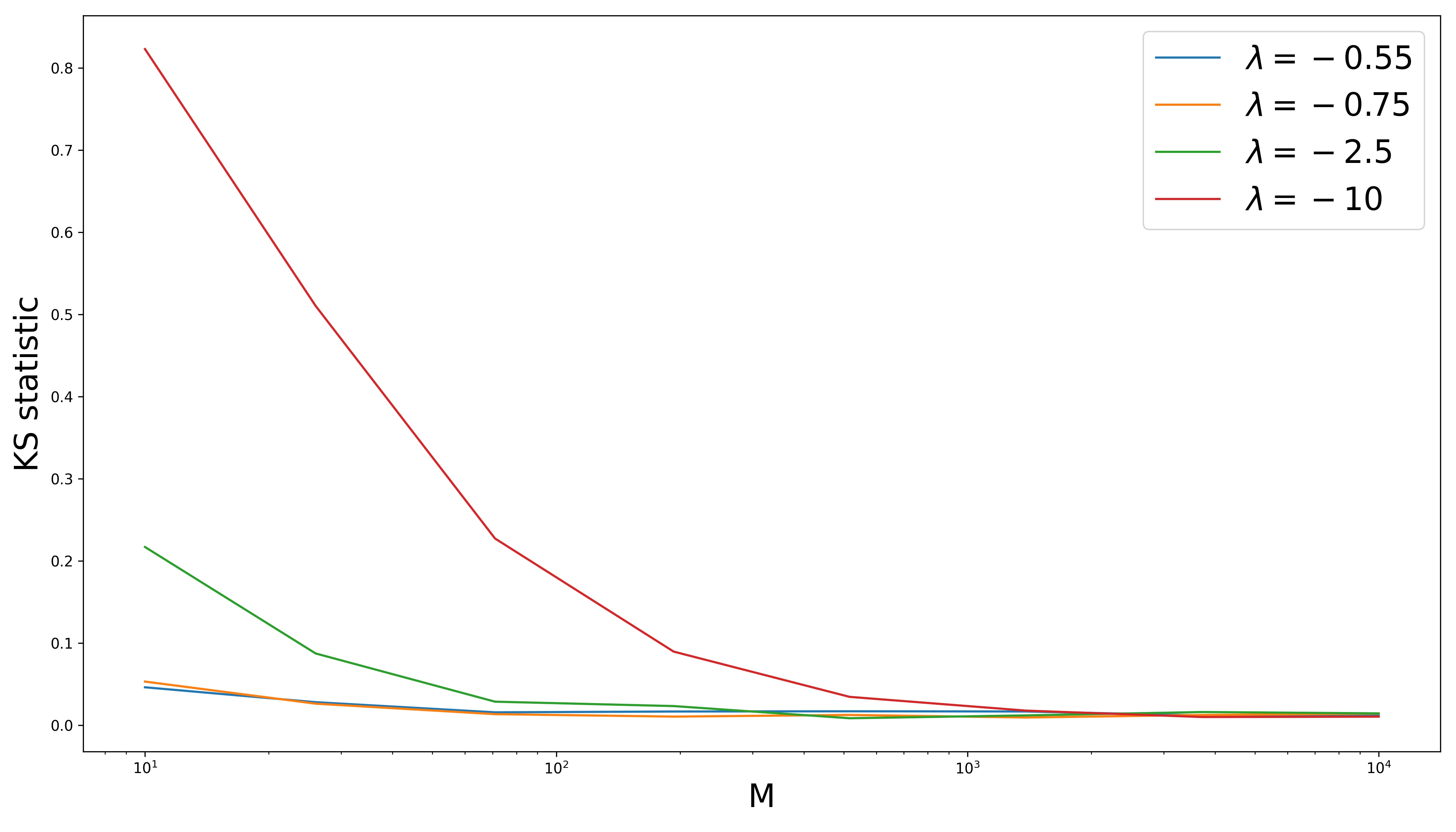}
\caption{KS statistics for $\lambda < -0.5$ and various series lengths $M$. }
\label{series_length_KS_test} 
\end{figure}

The results of such a KS test are reported in Fig. \ref{series_length_KS_test}, for values of $|\lambda|<-0.5$, varying shot noise series length $M$, and using random sample sizes of 10000. Remaining parameters were $\gamma=0.1$, $\delta=2$. It can be seen that the convergence of the KS statistic is slower as $\lambda$ becomes more negative, as alluded to in earlier analysis. The final p-values at $M=10000$ are typically well above 0.1, indicating that the hypothesis cannot be rejected for this $M$, at a confidence level of 0.1, with considerably earlier convergence for the smaller absolute values of $\lambda$.


For $0<|\lambda|<0.5$ we observed very rapid convergence of the KS statistics to acceptable levels as $M$ increases, typically within a few dozen terms in the series, so we do not display these results.  

\begin{figure}[tbp]
  \includegraphics[width=0.9\textwidth]{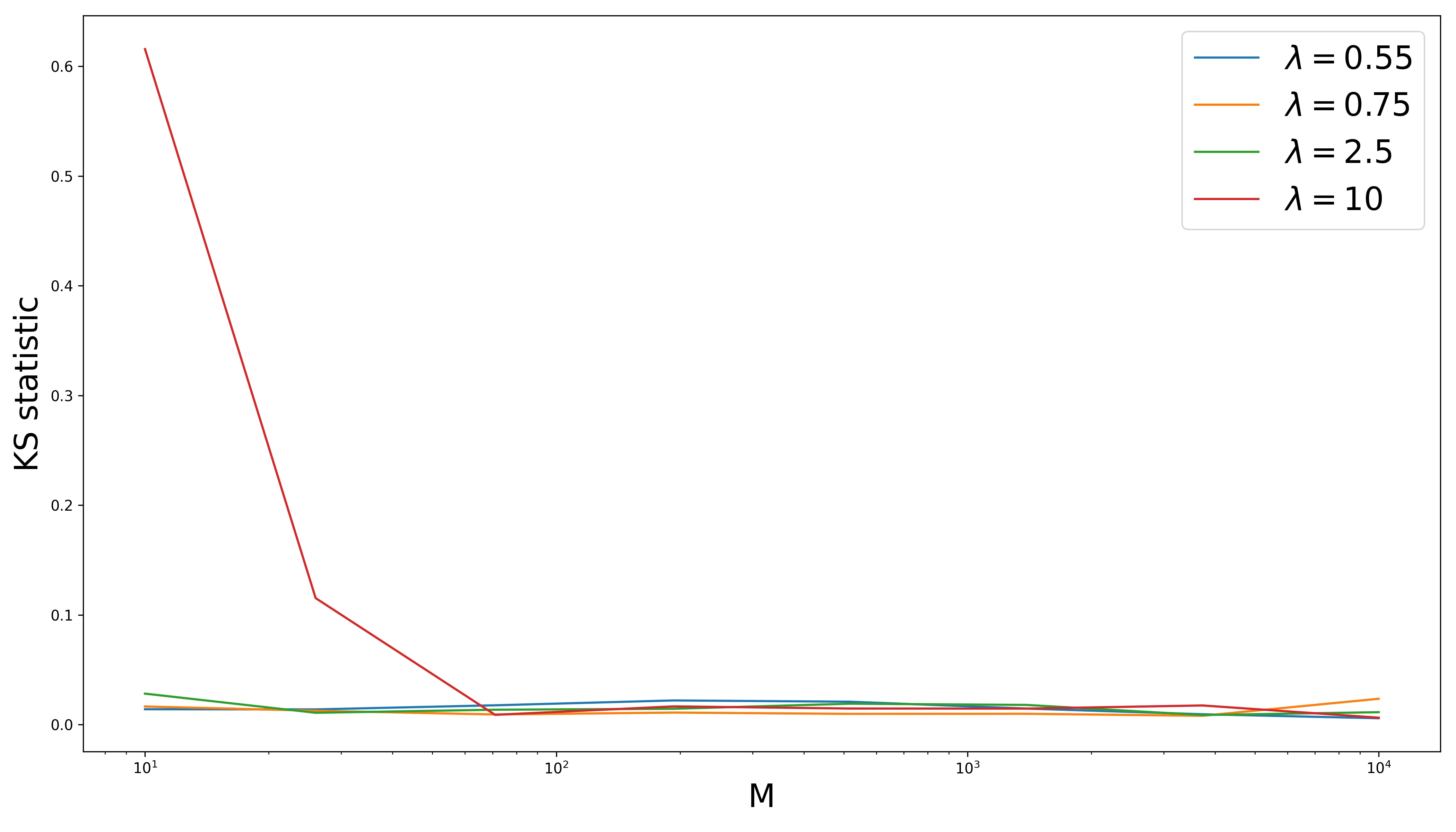}
\caption{KS statistics for $\lambda > 0.5$ and various series lengths $M$. }
\label{series_length_KS_test_pos} 
\end{figure}

For $\lambda>0.5$ Fig. \ref{series_length_KS_test_pos} gives the equivalent set of KS curves, showing more rapid convergence than their negative counterparts (presumably because of the additional gamma process that is mixed into the point process) and again showing p-values typically well above 0.1 for $M=10000$.

Of course, we do recognise that our generated GIG processes are approximate, and so it should always be possible to reject the null hypothesis for sufficiently large sample sets, but nevertheless these tests do give a helpful indication of the new algorithm's performance for different ranges of $\lambda$. Similar behaviours were observed for other settings of parameters $\gamma$ and $\delta$.

Finally the computational burden of the algorithms are linear in $M$, the number of terms in the series. Moreover there are no random waiting times since the accept-reject steps are performed as one-off tests for each point generated in the series. The most  significant contributors to the load are the calculation of the various acceptance probabilities and the sampling of the auxiliary variables $z$. The Matlab  and Python code runs at similar speed on standard laptop platforms. On a Microsoft Surface (Intel(R) Core(TM) i7-1065G7 CPU @ 1.30GHz   1.50 GHz) the time for computation was very roughly $10^{-6}M$s per random process realisation with $|\lambda|\geq 0.5$ and approximately double that for the more complex $|\lambda|<0.5$ case. Note also that in Algorithm \ref{Q_GIG_gen} some significant parallelisation is possible, since Steps 1.-3. do not depend on $\lambda$. Thus it would be possible to simulate many different realisations for different $\lambda$ by performing one set of steps 1.-3. and then performing step 4. independently for each $\lambda$ value required. This could be of significant value in ensemble and Monte Carlo inference procedures, as well as for rapid visualisation of different $\lambda$ scenarios.

\section{Discussion}
This paper has presented a generic simulation methodology for GIG processes. The methods are simple and efficient to implement and have good acceptance rates, which will make them of use in applications for practitioners. Moreover, we provide code in Matlab and Python in order to give researchers immediate access to the methods. 

Simulation of GIG processes opens up the possibility of simulation and inference for many more complex processes; in particular, a direct extension takes the sampled GIG process and generates other processes within the generalised shot-noise methodology \cite{Rosinski_2001}:
\[
W(t)=\sum_{i=1}^{\infty} H(U_{i},\Gamma_{i}){\cal I}(V_{i}\leq t)
\]
where $U_i$ are iid uniform random variates and $H(u,\gamma)$ a non-increasing function of $\gamma$.
Of particular interest will be the mean-variance mixture of Gaussians, with:
\[
H(U,\gamma)\sim {\cal N}(\mu_{W} h(\gamma),\sigma_{W}^{2} h(\gamma))
\]
and $J_{i}=h(\Gamma_{i})$ is the $i$th ordered jump of the simulated GIG process.
This approach, which is an exactly equivalent process to the time-changed Brownian motion description of \cite{B_N_1997}, leads directly to a simulation method for the generalised hyperbolic (GH) process, and its conditionally Gaussian form will enable inference for these processes, using a conditionally Gaussian approach similar in spirit to \cite{Lemke_Godsill_2015,Godsill_Riabiz_Kont_2019}. 
Our continued work on these processes will study these inferential approaches with the GIG/GH model and also their use as driving processes for stochastic differential equations, again extending the approach of \cite{Godsill_Riabiz_Kont_2019} to these more general classes of process.

\printbibliography

\end{document}